\documentclass[pre,aps,showpacs,twocolumn,longbibliography]{revtex4-1}

\usepackage{graphicx}
\usepackage{amsfonts}
\usepackage{amssymb}
\usepackage{amsmath}
\usepackage{float}
\usepackage{color}

\begin{document}

\title{Ensemble Theory for Stealthy Hyperuniform Disordered Ground States}

\author{S. Torquato}

\email{torquato@electron.princeton.edu}

\affiliation{Department of Chemistry, Department of Physics,
Princeton Institute for the Science and Technology of
Materials, and  Program in Applied and Computational Mathematics,
Princeton University,
Princeton, New Jersey 08544, USA}

\author{G. Zhang}

\affiliation{Department of Chemistry, Princeton University,
Princeton, New Jersey 08544, USA}

\author{F. H. Stillinger}

\affiliation{ Department of Chemistry, Princeton University,
Princeton, New Jersey 08544, USA}

\date{\today}

\begin{abstract}

It has been shown numerically that systems of particles interacting with isotropic
``stealthy" bounded long-ranged  pair potentials (similar to Friedel oscillations)
have classical ground states that are (counterintuitively) disordered, hyperuniform, and highly degenerate.
Disordered hyperuniform systems have received attention recently
because they are distinguishable exotic states of matter poised between a crystal
and liquid that are endowed with novel thermodynamic and physical properties.
The task of formulating an ensemble theory that yields
analytical predictions for the structural characteristics and other properties of stealthy degenerate
ground states in $d$-dimensional Euclidean space $\mathbb{R}^d$ is highly nontrivial because the dimensionality of the  configuration
space depends on the number density $\rho$ and there is a multitude of ways
of sampling the ground-state manifold, each with its own probability measure
for finding a particular ground-state configuration. The purpose 
of this paper is to take some initial steps in this direction.
Specifically,  we derive general exact relations for thermodynamic properties
(energy, pressure, and isothermal compressibility)
that apply to any ground-state ensemble as a function of $\rho$ in any $d$, and we show   how  disordered degenerate ground states arise as part of the ground-state manifold. 
We also derive  exact integral conditions that  both the pair correlation function 
$g_2(r)$ and structure factor $S(k)$ must obey for any $d$. We then specialize our results to the canonical ensemble 
(in the zero-temperature limit) by exploiting an ansatz that stealthy states behave remarkably like 
``pseudo" equilibrium  hard-sphere systems in Fourier space.
Our theoretical predictions for $g_2(r)$ and $S(k)$ are in excellent
agreement with computer simulations across the first three space dimensions. 
These results are used to obtain  order metrics, local number variance and nearest-neighbor
functions across dimensions.  We also derive accurate analytical formulas for the structure factor 
and thermal expansion coefficient for the excited states at sufficiently small temperatures for any $d$.
The development of this theory provides new insights regarding 
our fundamental understanding of the nature and formation of
low-temperature states of amorphous matter. Our work also offers challenges to experimentalists
to synthesize stealthy ground states at the molecular level.

\end{abstract}

\maketitle

\section{Introduction}

The equilibrium structure  and phase behavior  of soft matter systems span from the relatively simple,
as found in  strongly repulsive colloidal particles, to  the highly complex, as seen in microemulsions and polymers
\cite{St76,Fr84,Chaik95,Gu97,Cr00,Chaik05,Ml06,Za08,Ca12,Ha14}.
Soft matter has been fruitfully microscopically modeled 
as classical many-particle systems in which the particles (or metaparticles) interact
with effective pair potentials. Bounded (soft) effective interactions have
been particularly useful in modeling polymer systems, and they display zero-temperature ground states 
with a rich variety of crystalline structures, depending on the composition of the constituents
and interaction parameters \cite{St76,Gu97,Ml06,Za08,Ca12,Ha14}.

We have previously used a ``collective-coordinate" approach to generate numerically 
exotic classical  ground states of many particles 
interacting with certain bounded isotropic long-ranged pair potentials 
in one-, two- and three-dimensional  Euclidean space dimensions \cite{Fa91,Uc04b,Uc06b,Ba08,Ba09a,Ba09b,Za11b}
as well as with anisotropic potentials \cite{Mar13}.
It was shown that the constructed ground states across dimensions  are the expected crystal
structures in a low-density regime \cite{Fa91,Uc04b,Su05,Ba08}, but 
above some critical density, there is a phase transition to
ground states that are, counterintuitively, disordered  (statistically isotropic with no long-range order), hyperuniform, and highly degenerate \footnote{Indeed, the number of ground-state degeneracies is uncountably infinite for a finite number
of particles, which distinguishes it from disordered ground states
found in classical Ising-like spin systems in which the number of degeneracies
is finite for a finite number of spins  \cite{He10,Bal10,Di13,Ma13d}. We note in passing that
while quantum spin liquids have disordered ground states, they are effectively  unique \cite{Bal10}.}.
These unusual amorphous states of matter have been shown to be endowed with novel thermodynamic and physical properties \cite{Ba08,Ba09a,Ba09b,Fl09b,Fl13,Man13b} and belong to the more general class of disordered ``hyperuniform" systems, which have been attracting attention recently, as detailed below.

The disordered ground states are highly degenerate with a configurational
dimensionality that depends on the density, and there are an infinite number of 
distinct ways to sample this complex ground-state manifold,  each with its own 
probability measure. For these reasons, it is theoretically very challenging to devise  ensemble
theories that are  capable of predicting structural attributes 
and other properties of the ground-state configurations. A new type of statistical-mechanical
theory must be invented to characterize these exotic states of matter. The purpose 
of this paper is to take some initial steps in this direction.
However, to motivate the theoretical formalism, it is instructive to first
briefly review the collective-coordinate numerical procedure that we have used to achieve
disordered ground states.

In the simplest setting, we previously examined pairwise additive potentials 
$v({\bf r})$ that are bounded and integrable such that their
Fourier transforms ${\tilde v}({\bf k})$ exist. If $N$
identical point particles reside in a fundamental region $F$ of volume $v_F$ in $d$-dimensional
Euclidean space $\mathbb{R}^d$ 
at positions ${\bf r}^N \equiv {\bf r}_1, \ldots, {\bf r}_N$ under periodic
boundary conditions, the total potential energy can be expressed
in terms of ${\tilde v}({\bf k})$ as follows:
\begin{equation}
\Phi({\bf r}^N) =\frac{1}{2v_F} \left[\sum_{\bf k} {\tilde v}({\bf k})|{\tilde n}({\bf k})|^2
-N \sum_{\bf k} {\tilde v}({\bf k})\right],
\label{pot}
\end{equation}
where ${\tilde n}({\bf k }) = \sum_{j=1}^{N} \exp(-i{\bf k \cdot r}_j)$ is
the complex collective density variable, which can be viewed as a nonlinear transformation
from the finite set of particle coordinates ${\bf r}_1, \ldots, {\bf r}_N$
to the complex functions ${\tilde n}({\bf k })$ that depend on the
infinite set of wave vectors $\bf k$ in reciprocal space appropriate to the
fundamental cell $F$. The crucial idea is that if ${\tilde v}({\bf k})$
is defined to be bounded and positive with support in the radial
interval $0 \le |{\bf k}| \le K$ and if the particles are arranged so 
that $|{\tilde n}({\bf k})|^2$, a quantity proportional to the structure factor $S({\bf k})$,  
is driven to its minimum value of zero for all wave vectors where ${\tilde v}({\bf k})$ has support (except
$\bf k=0$), then it is clear from relation (\ref{pot})
that the system must be at its ground state or global energy minimum.
We have referred to these ground-state configurations
as ``stealthy" \cite{Ba08} because the structure factor  $S({\bf k})$ (scattering pattern)
is zero for $0 < |{\bf k}| \le K$, meaning that they completely suppress
single scattering of incident radiation for these wave vectors and, thus, are
transparent at the corresponding wavelengths \footnote{More generally, stealthy configurations can be those ground states that
correspond to minimizing $S(k)$ to be zero at other sets of wave vectors, not
necessarily in a connected set around the origin, specific examples of which were investigated
in Ref. \cite{Ba08}. We have also used the collective-coordinate technique to target more general forms
of the structure factor for a prescribed set of wave vectors such that $S({\bf k})$
is not minimized to be zero in this set (e.g., power-law forms and positive constants) \cite{Uc06b,Ba08,Za11b}. There the resulting configurations
are the ground states of interacting many-particle systems with 2-,3- and  4-body interactions.}.
Various optimization techniques were employed
to find the globally energy-minimizing configurations  within an exceedingly small numerical 
tolerance \cite{Fa91,Uc04b,Uc06b,Ba08,Ba09a,Ba09b,Za11b,Mar13}. Generally, a numerically obtained 
ground-state configuration depends on the number of particles $N$ within the fundamental cell, initial particle configuration,
shape of the fundamental cell,  and  particular optimization technique employed.

As the number of $\bf k$ vectors for which $|{\tilde n}({\bf k})|$ is constrained to be zero increases, i.e., as $K$ increases,
the dimensionality of the ground-state configuration manifold $D_C$ decreases. 
 Because  $|{\tilde n}({\bf k})|$ is inversion symmetric, the
number of wave vectors contained in a sphere of radius $K$ centered at the origin  must be an odd integer,
say $2M(K)+1$, and thus $M(K)$ is the number of  independently
constrained wave vectors \footnote{Since both the real and imaginary part
contributions to  $|{\tilde n}({\bf k})|$ are zero for each wave vector $\bf k$ in the constrained region or \textit{exclusion
zone}, the total number of independent constrained degrees of freedom is $2M(K)$. 
Hence, in the large-system limit, $\chi=1/2$ 
is the critical value when there are no longer any degrees of freedom 
that can be independently constrained to be zero (not $\chi=1$), according to this simple counting argument. 
The reason why we use the definition (\ref{chi})
is that in the more general case when $|{\tilde n}({\bf k})|$ is constrained to be positive (not zero) for some
set of wave vectors \cite{Uc06b,Ba08,Za11b}, $\chi=1$  is indeed the critical value when one runs out of degrees
of freedom that can be independently constrained in the large-system limit.}.  The parameter 
\begin{equation}
\chi = \frac{M(K)}{d(N-1)},
\label{chi}
\end{equation}
which is inversely proportional to density, gives a measure of the relative fraction of constrained degrees of freedom compared
%DIF > represents the ratio of the number of constrained degrees of freedom in the system 
to the total number of degrees of freedom $d(N-1)$
(subtracting out the system  translational degrees of freedom). We show
in Sec. \ref{dim} that the dimensionality of the configuration space per particle is given
by $d(1-2\chi)$ in the thermodynamic limit.

\begin{figure}[H]
\begin{center}
\includegraphics[  width=2.75in, keepaspectratio,clip=]{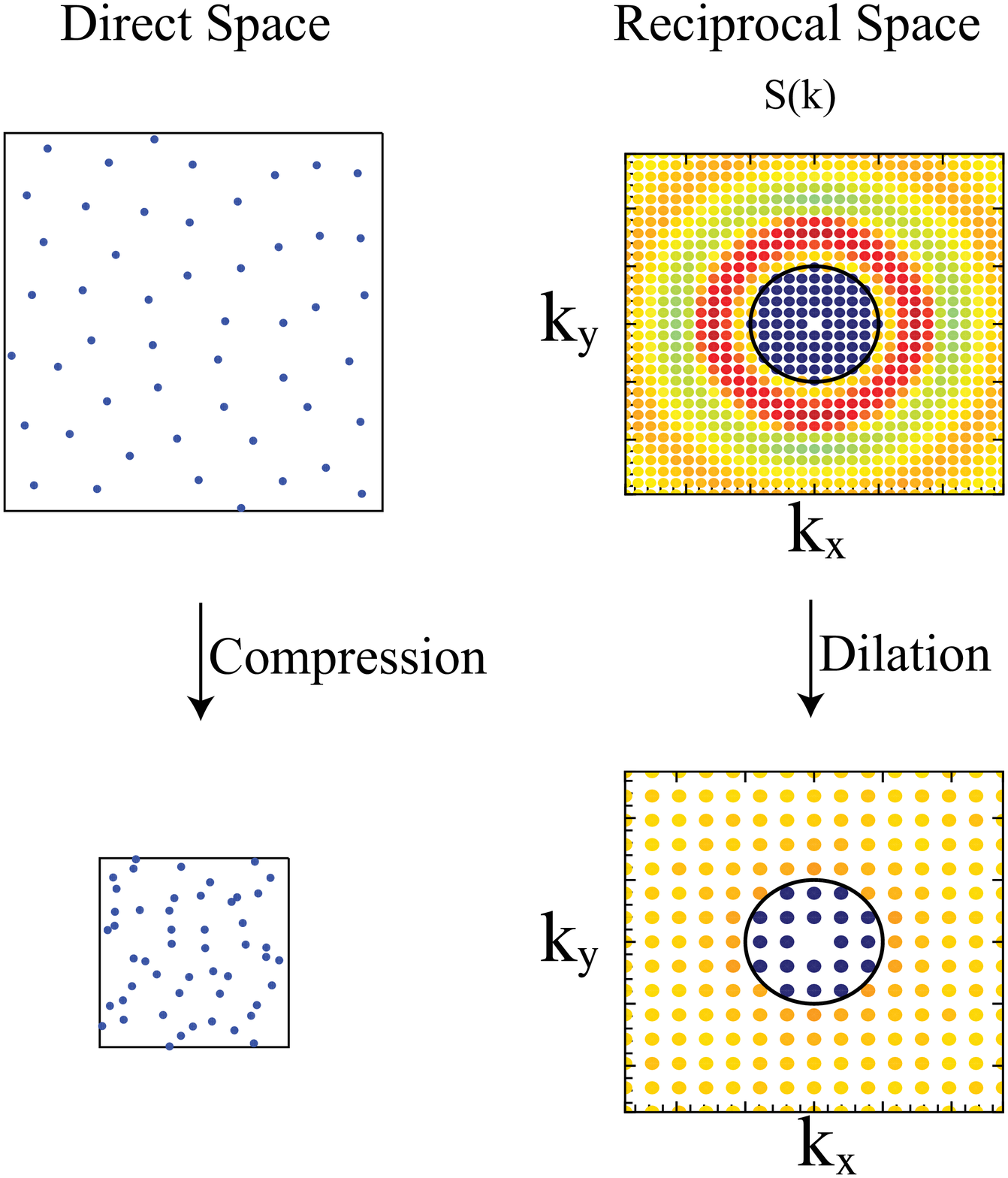}\vspace{-0.1in}
\caption{Schematic illustrating the inverse relationship between the
direct-space number density $\rho$ and relative fraction of constrained degrees of freedom $\chi$ 
for  a fixed reciprocal-space exclusion-sphere radius $K$ (where dark blue  $\bf k$ points signify zero intensity
with green, yellow, and red points indicating increasingly larger intensities) 
for a stealthy ground state. A compression of a disordered ground-state configuration {with a fixed number of particles $N$ }in direct space leads to a dilation of the lattice spacing in reciprocal space. 
This means that during the compression process, the $\bf k$ points for which $|{\tilde n}({\bf k})|$ is zero
associated with the initial uncompressed system move out of the exclusion zone; i.e., the value of $M(K)$
[cf. Eq. (\ref{chi})] decreases. Since  there are fewer constrained degrees of freedom (dimensionality
of the ground-state configuration manifold increases), the disordered direct-space configuration becomes 
less spatially correlated. {For a fixed $N$ in the limit $\rho \rightarrow \infty$ (i.e. system volume $v_F \to 0$), }every
${\bf k}$ point  (except the origin) is expelled from the exclusion zone, and the system tends to an ideal-gas configuration
\cite{X},
even if it is not an ideal gas thermodynamically, as shown in Sec. \ref{dim}. }

\label{rho_chi}
\end{center}
\end{figure}
\vspace{-0.28in}

It is straightforward to see why,  for sufficiently small $\chi$, ground states exist that are highly degenerate 
and typically disordered for sufficiently large $N$; see Fig. \ref{rho_chi}.
%which can be regarded as classical many-particle
%analogs of quantum spin liquids \cite{An73,Ya11}.
Clearly, when the system is free of any constraints, i.e., if $\chi=0$, it is a noninteracting classical ideal gas. While it is unusual to think of classical ideal-gas configurations as  ground states, at $T=0$,
they indeed are global energy-minimizing states that are highly degenerate and typically disordered for large enough $N$.
While the ground-state manifold contains periodic configurations (e.g., Bravais lattices and lattices
with a basis), these are sets of zero measure in the thermodynamic limit. Clearly,
 if  $\chi$ is made positive but very small, the ground states remain disordered
and highly degenerate, even if the dimensionality of the configuration space $D_C$ has now been suddenly reduced 
due to the imposed constrained degrees of freedom, the number of which is determined by the radius $K$.
From relation (\ref{chi}), we see that if $K$ is fixed, 
  {configurations with ideal-gas-like pair correlation functions } correspond to the limit $\chi \rightarrow 0$
or, equivalently, to the limit $\rho \rightarrow \infty$ \cite{X}. The latter situation runs counter
to traditional understanding that  ideal-gas configurations correspond to the opposite zero-density limit
of classical systems of particles. The reason for this inversion of limits
is due to the fact that a compression of the system in direct space
leads to a dilation of the lattice spacing in reciprocal space, as
schematically shown in Fig. \ref{rho_chi}. While it is not surprising that 
the configuration space is fully connected for sufficiently small $\chi$,
quantifying its topology as a function of $\chi$ for all allowable $\chi$ is an outstanding problem, which is
discussed further in the Conclusions.

\begin{figure}[H]
\begin{center}
{\includegraphics[  width=1.6in, keepaspectratio,clip=]{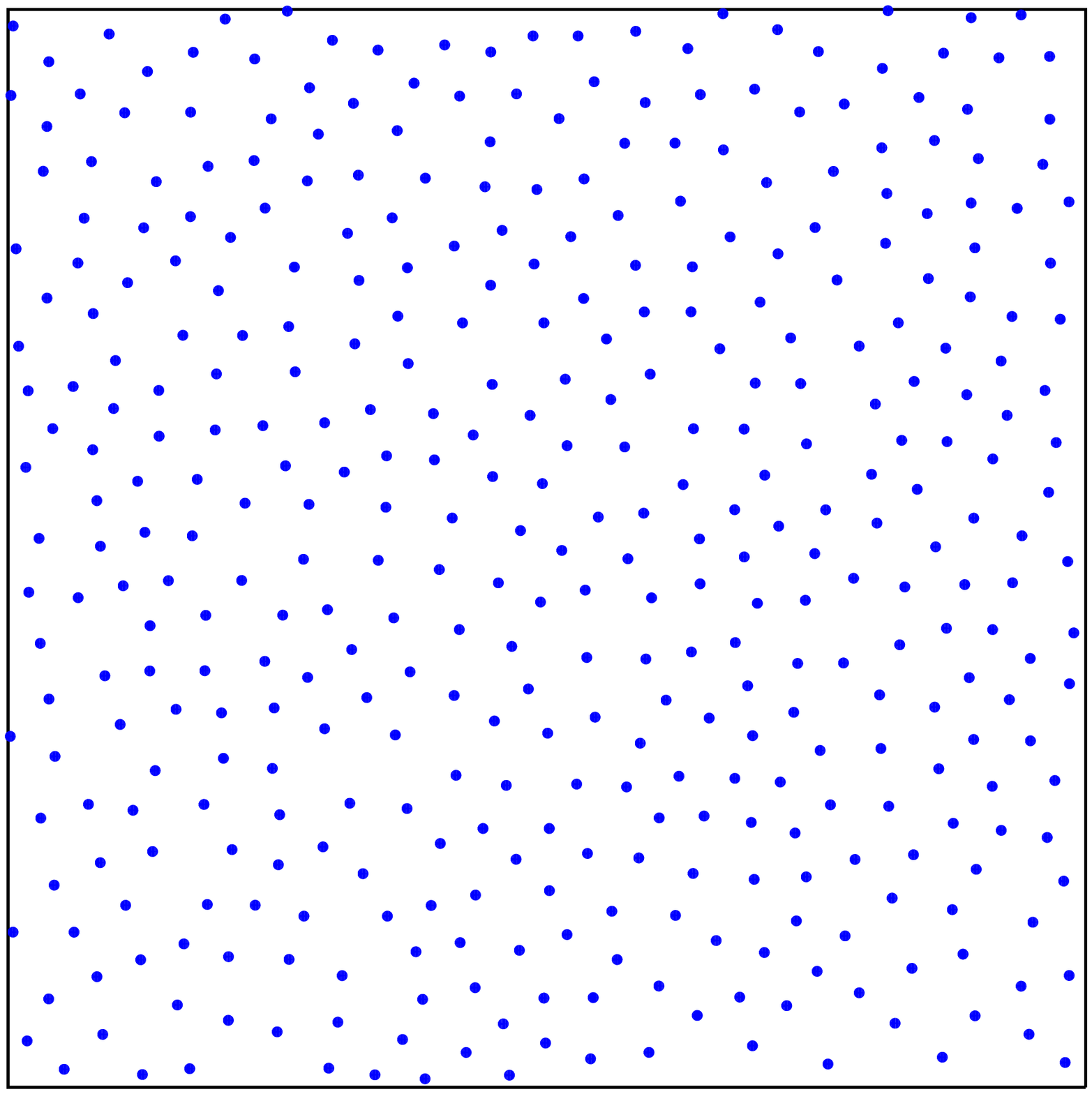} \includegraphics[  width=1.6in, keepaspectratio,clip=]{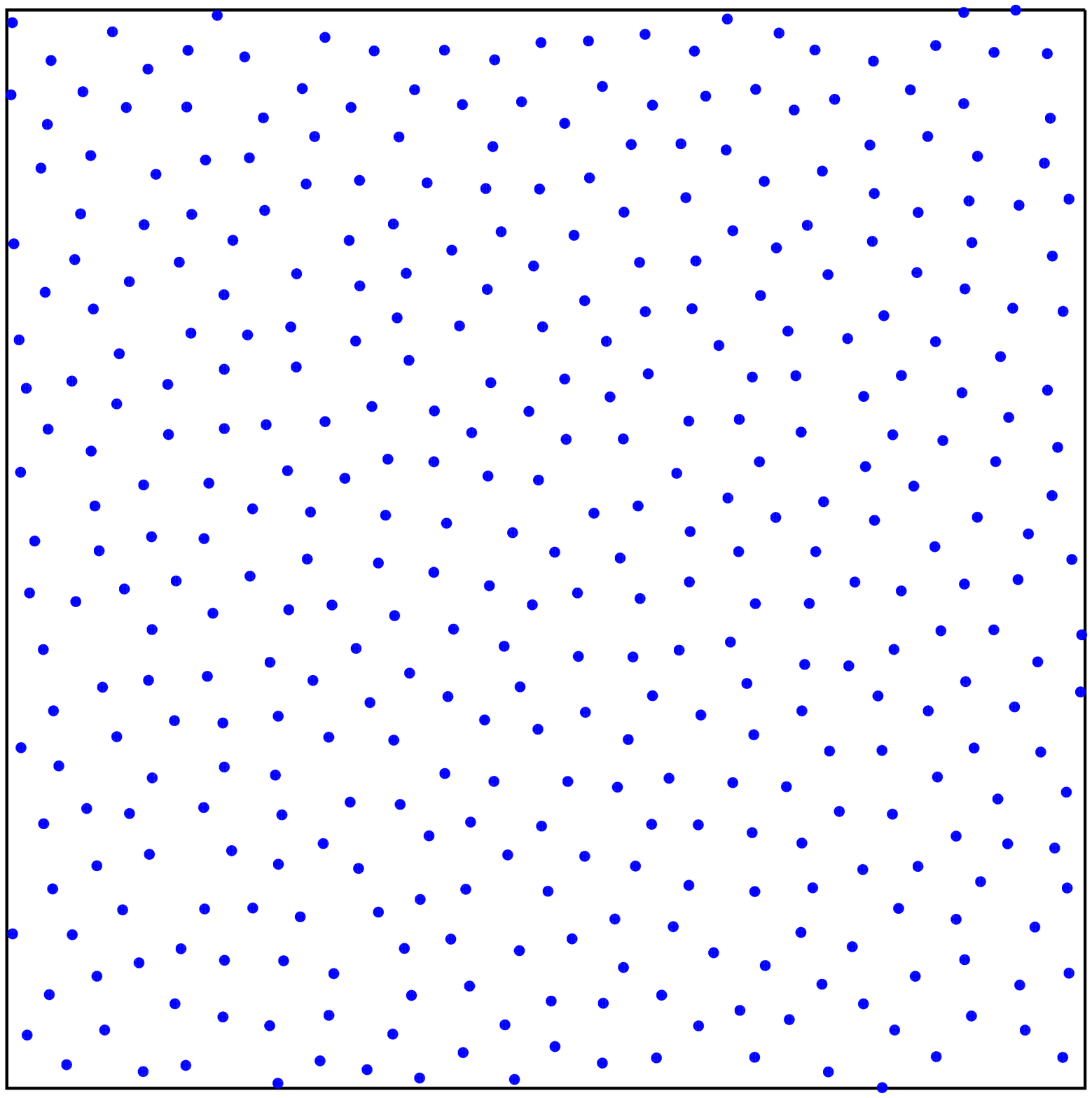}}
%\vspace{-0.3in}
\caption{A disordered {\it nonhyperuniform} configuration (left panel)
and a disordered {\it hyperuniform} configuration (right panel). We arrive at the configuration on the right by very small collective displacements of the particles on the left
via the methods described in Ref. \cite{Uc04b}.
(Each particle on average moves a root-mean-square distance that is about an order of magnitude smaller than 
the mean-nearest-neighbor distance as measured by the configuration proximity metric \cite{Ba11b}.)
These two examples show that it can be very difficult to detect hyperuniformity by eye,
and yet their large-scale density fluctuations are dramatically different (``hidden order").}
\label{stealthy}
\end{center}
\end{figure}

It is noteworthy that stealthy point patterns (disordered or not) 
constitute a special class of so-called hyperuniform states of matter.
Hyperuniform systems are characterized by vanishing (normalized) density fluctuations at large length scales; i.e.,
the structure factor $S({\bf k})$ tends to zero in the limit $|{\bf  k}| \rightarrow 0$ ~\cite{To03a} (see
Sec. \ref{hyp} for details). The hyperuniformity concept provides a means of categorizing crystals, quasicrystals, and special disordered systems according to the degree to which large-scale density fluctuations are suppressed ~\cite{To03a,Za09}. Disordered hyperuniform patterns, of which disordered stealthy systems are special cases, behave more like crystals in the manner they suppress large-scale density fluctuations, and yet they also resemble typical
statistically isotropic liquids and glasses with no Bragg peaks. 
In this sense, they have a ``hidden order" on large length scales that is not apparent
at small length scales, even if short-range order is present; see Fig. \ref{stealthy} for a vivid
illustration. During the last decade, a variety of disordered hyperuniform states have been identified  that exist as both equilibrium and nonequilibrium phases,
including maximally random jammed particle packings \cite{Do05d,Za11a,Ji11c,Ch14a},  jammed athermal granular media~\cite{Be11}, jammed thermal colloidal packings~\cite{Ku11,Dr15}, cold atoms~\cite{Le14}, transitions in nonequilibrium systems \cite{He14,Ja15},
surface-enhanced Raman spectroscopy \cite{Zi15},
terahertz quantum cascade laser  \cite{De15}, wave dynamics in disordered potentials based on supersymmetry \cite{Yu15},
avian photoreceptor patterns \cite{Ji14}, and certain Coulombic systems~\cite{To08c}. Moreover, disordered hyperuniform materials possess novel physical properties potentially important for applications in photonics~\cite{Fl09b,Fl13,Man13b,Ha13,La14} and electronics \cite{He13,Xie13,Be15}.

The well-known compressibility relation from statistical mechanics \cite{Ha86} provides 
some insights about the relationship between temperature $T$ and hyperuniformity
for equilibrium  systems at number density $\rho$:
\begin{equation}
S(k=0)=\rho k_B T \kappa_T.
\label{comp}
\end{equation}
We see that any ground state ($T=0)$ in which the isothermal
compressibility $\kappa_T$ is bounded and positive must be hyperuniform
because the structure factor $S(k =0)$ must be zero. This includes crystals as well
as exotic disordered ground states such as stealthy ones. However, in order to have a hyperuniform
system at positive $T$, the isothermal compressibility must be zero; i.e.,
the system must be incompressible \cite{Za11b} (see Refs. \cite{To03a} and \cite{To08c}
for some examples). Subsequently, we will use relation (\ref{comp})
to draw some conclusions about the excited states associated with stealthy ground states.

Our general objective is the formulation of a predictive ensemble theory for the 
thermodynamic and structural properties of stealthy degenerate disordered ground states in arbitrary space dimension $d$
that complements previous numerical work on this topic \cite{Fa91,Uc04b,Uc06b,Ba08,Ba09a,Ba09b}.
After providing basic definitions and describing a family of isotropic stealthy potentials
(Secs. \ref{def} and \ref{family}), we derive general exact relations for thermodynamic properties
(energy, pressure, and isothermal compressibility)
that apply to any well-defined ground-state ensemble as a function of the number
density or, equivalently, $\chi$ in any space dimension $d$ (Sec. \ref{ensemble}) . We subsequently
derive some exact integral conditions that  both the pair correlation function $g_2(r)$
and structure factor $S(k)$ must obey (Sec. \ref{pair}). The existence of periodic stealthy ground states
enables us to show  how  disordered degenerate ground states arise as part of the ground-state manifold
for sufficiently small $\chi$ (Sec. \ref{exist}). Subsequently, we derive analytical formulas for the pair statistics
[$g_2(r)$ and $S(k)$] for sufficiently small $\chi$  in the canonical ensemble in the limit 
that the temperature $T$ tends to zero (Sec. \ref{pseudo})
by exploiting an ansatz that stealthy states behave like ``pseudo" equilibrium  hard-sphere systems in Fourier space.
Our theoretical predictions for $g_2(r)$ and $S(k)$ are in excellent
agreement with computer simulations across the first three space dimensions. These results are then used to 
predict, with high accuracy, other structural characteristics of stealthy ground states across dimensions, such as order metrics, local number variance, and nearest-neighbor
functions (Secs. \ref{var} and \ref{near}).
Subsequently, we derive analytical formulas for the structure factor 
and thermal expansion coefficient for the associated excited states for sufficiently small temperatures
(Sec. \ref{excite}). Finally, we provide concluding remarks in Sec. \ref{concl}.

\section{Definitions and Preliminaries}
\label{def}

Roughly speaking, a point process in $d$-dimensional Euclidean space $\mathbb{R}^d$ is a 
distribution of an  infinite number of points in $\mathbb{R}^d$ with the configuration ${\bf r}_1,{\bf r}_2,\ldots$
at a well-defined number density  (number of points per unit volume). 
For a statistically homogeneous point process in $\mathbb{R}^d$
at number density $\rho$ \cite{Note5},  the quantity
$\rho^n g_{n}({\bf r}^n)$  is
the probability density associated with simultaneously finding $n$ points at
locations ${\bf r}^n \equiv {\bf r}_1,{\bf r}_2,\dots,{\bf r}_n$ in $\mathbb{R}^d$ \cite{Ha86}.
 With this convention, each {\it $n$-particle correlation function} $g_n$ approaches
unity when all of the points become widely separated from one another.
Statistical homogeneity implies that $g_n$ is translationally
invariant and hence only depends on the relative displacements
of the positions with respect to any chosen system origin, {\it e.g.},
$g_n=g_n({\bf r}_{12}, {\bf r}_{13}, \ldots, {\bf r}_{1n})$,
where ${\bf r}_{ij}={\bf r}_j - {\bf r}_i$.

The pair correlation function $g_2({\bf r})$ is a particularly
important quantity. If the point process is also rotationally invariant (statistically
isotropic), then $g_2$ depends on the radial distance $r \equiv |{\bf r}|$ 
only, {\it i.e.}, $g_2({\bf r}) = g_2(r)$. Thus, it follows that the expected 
number of points $Z(R)$ found in a sphere of radius $R$ around a randomly
chosen point of the point process, called the {\it cumulative
coordination function}, is given by
\begin{equation}
Z(R)=\rho s_1(1) \int_0^R  x^{d-1} g_2(x) dx,
\label{cum}
\end{equation}
where $s_1(r)  =  2\pi^{d/2}r^{d-1}/\Gamma(d/2)$
is the surface area of a $d$-dimensional sphere of radius $r$.
The total correlation function $h({\bf r})$ is trivially related to $g_2({\bf r})$
as follows:
\begin{equation}
h({\bf r})\equiv g_2({\bf r})-1.
\label{tot}
\end{equation}
When there are no long-range correlations in the system, $h({\bf r}) \rightarrow 0$
or, equivalently, $g_2({\bf r}) \rightarrow 1$ as $|\bf r| \rightarrow \infty$.
The structure factor $S(\bf k)$, which plays a prominent role in this paper, is related to the Fourier
transform of $h(\bf r)$, denoted by ${\tilde h}({\bf k})$, via the expression
\begin{equation}
S({\bf k})\equiv 1+\rho {\tilde h}({\bf k}).
\label{factor}
\end{equation}

A lattice $\Lambda$ in $\mathbb{R}^d$ is a subgroup
consisting of integer linear combinations of vectors that constitute a basis for $\mathbb{R}^d$,
and thus, it represents  a special subset of point processes.
Here, the space  can be geometrically divided into identical regions $F$ called {fundamental cells}, each of which contains
just one point  specified by the lattice vector
\begin{equation}
{\bf p}= n_1 {\bf a}_1+ n_2 {\bf a}_2+ \cdots + n_{d-1} {\bf a}_{d-1}+n_d {\bf a}_d,
\end{equation}
where ${\bf a}_i$ are the basis vectors for a fundamental cell
and $n_i$ spans all the integers for $i=1,2,\cdots, d$. We denote by
$v_F$  the volume of $F$.  A lattice is 
called  a Bravais lattice in the physical sciences. Unless otherwise stated, we will
use the term lattice. Every lattice has a dual (or reciprocal) lattice $\Lambda^*$
in which the lattice sites are specified by the dual (reciprocal) lattice vector
 ${\bf q}\cdot {\bf p}=2\pi m$  {for all $\mathbf p$} , where $m=0, \pm 1, \pm 2, \pm 3 \cdots$.
The dual fundamental cell $F^*$ has volume $v_{F^*}=(2\pi)^d/v_F$.
This implies that the number density $\rho_{\Lambda}$ of $\Lambda$
is related to the number  density $\rho_{\Lambda^*}$ of the dual lattice $\Lambda^*$
via the expression 
\begin{equation}
\rho_{\Lambda} \rho_{\Lambda^*}=1/(2\pi)^d.
\label{rho*}
\end{equation}
 {Some common $d$-dimensional lattices are mathematically defined in Appendix~\ref{app_lattices}.
} 

\begin{figure}
\centerline{\includegraphics*[  width=2in,keepaspectratio]{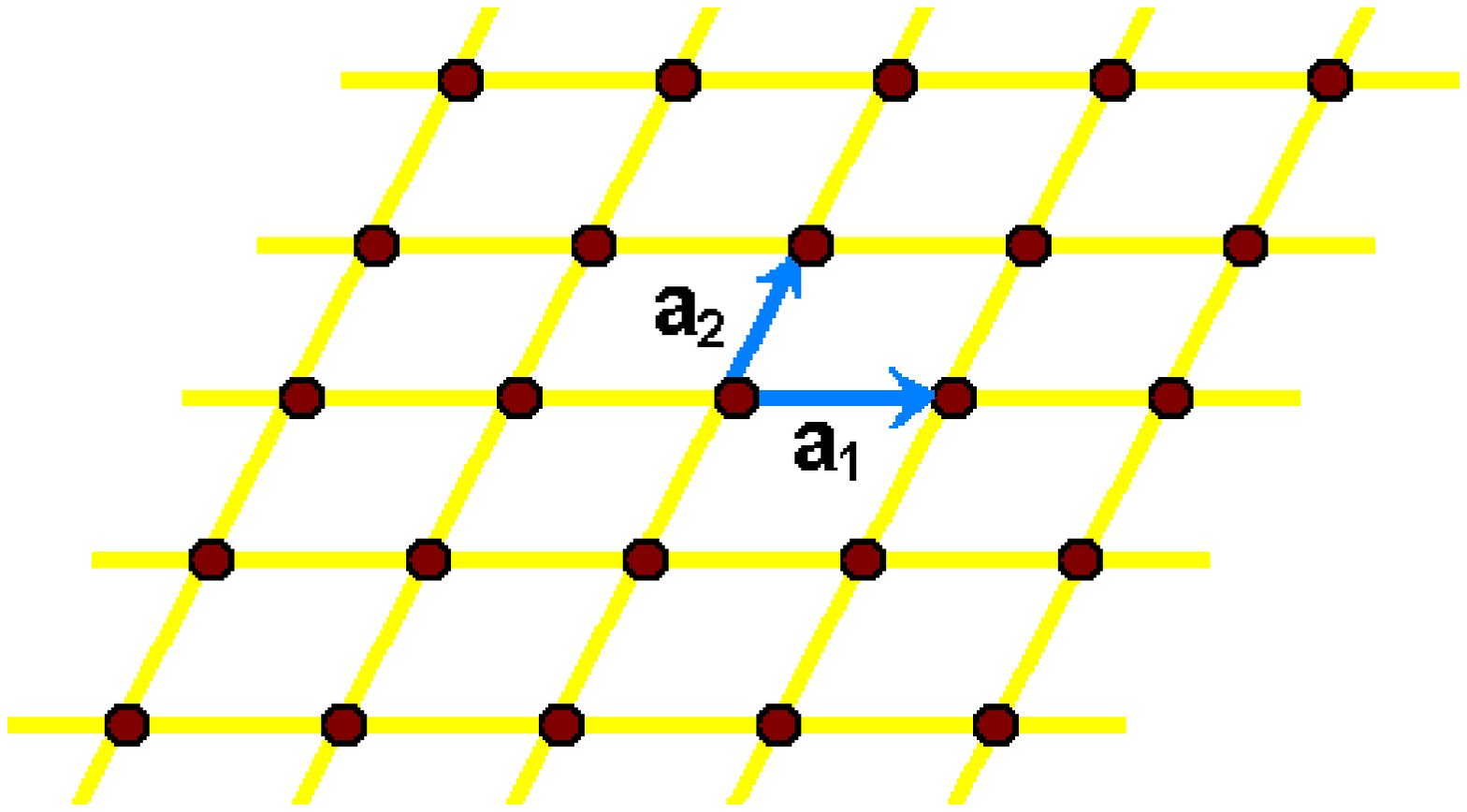} \hspace{-0.2in}
\includegraphics*[  width=2in,keepaspectratio]{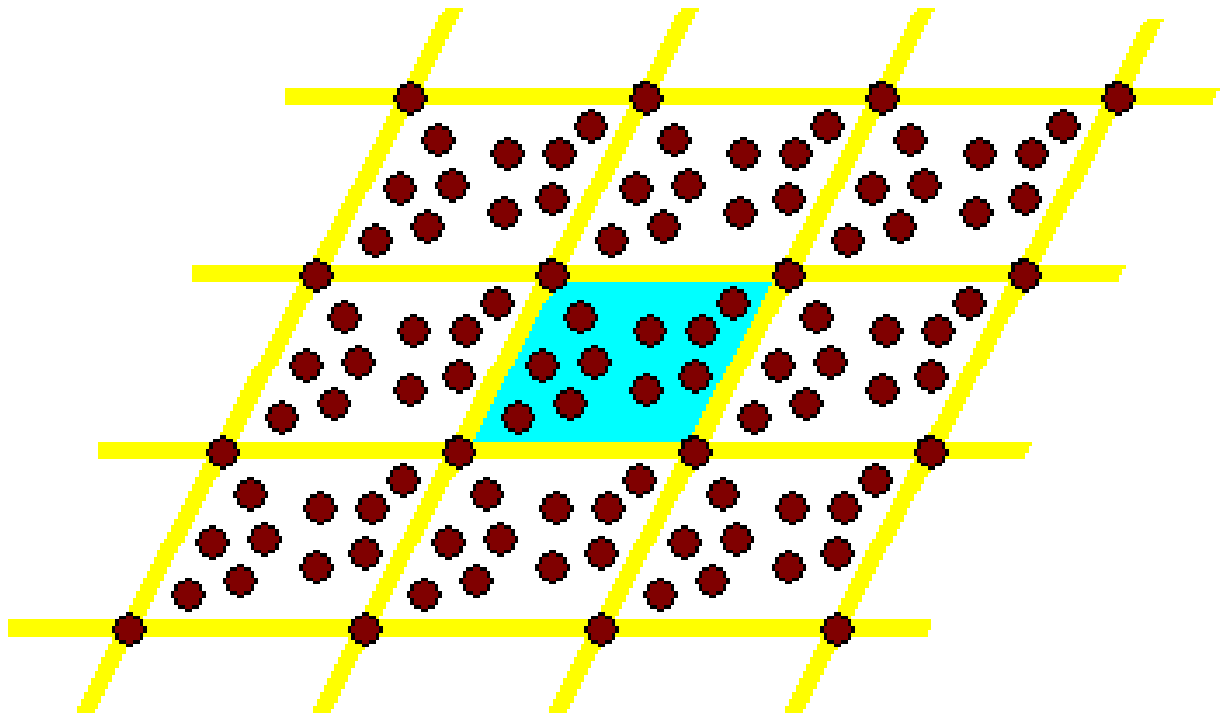}}
\caption{(Bravais) lattice with one particle per fundamental cell (left panel)
and a periodic crystal with multiple particles per fundamental cell (right panel).}
\label{lattice}
\end{figure}

A periodic point process (crystal) is a
more general notion than a lattice because it is
obtained by placing a fixed configuration of $N$ points (where $N\ge 1$)
within a fundamental cell $F$ of a lattice $\Lambda$, which
is then periodically replicated. Thus, the point process is still
periodic under translations by $\Lambda$, but the $N$ points can occur
anywhere in $F$; see Fig. \ref{lattice}.

\subsection{Hyperuniform point processes}
\label{hyp}

Consider uniformly sampling the number of points that are contained within a spherical window
of radius $R$ of a point process in $\mathbb{R}^d$. 
 A hyperuniform point process has the property that the 
local number variance $\sigma^2(R)$ grows more slowly than  $R^d$  \cite{To03a}. 
Because $\sigma^2(R)$ is exactly
related to a $d$-dimensional volume integral of the structure factor $S({\bf k})$  (see Sec. \ref{var}), this implies that hyperuniform states of matter possess
infinite-wavelength density fluctuations (appropriately normalized) that vanish; i.e., $S({\bf k})$
obeys the condition
\begin{equation}
\lim_{|{\bf k}| \rightarrow 0} S({\bf k}) \rightarrow 0,
\label{hyper}
\end{equation}
which means they are poised at an ``inverted" critical point with associated scaling exponents \cite{To03a}.
For a Poisson (spatially uncorrelated) point process and many disordered point patterns, including typical liquids and structural glasses, the number variance grows like the volume of the window, i.e., $\sigma^2(R) \sim R^d$, implying
that $S({\bf k})$ is positive at ${\bf k=0}$. All {\it perfect} crystals and quasicrystals are hyperuniform
such that $\sigma^2(R) \sim R^{d-1}$; in other words, the variance grows like
window surface area. By contrast, it is much more unusual to find disordered systems
that are also hyperuniform. In recent years, evidence has been emerging 
that disordered hyperuniform many-particle systems can be regarded as
new distinguishable states of disordered matter (see examples given in the Introduction). Whenever the structure factor
goes to zero with the power-law form $S({\bf k}) \sim |{\bf k}|^\alpha$, the number variance
has the following large-$R$ asymptotic scaling that depends on the value of the exponent $\alpha$ \cite{Za09,Za11b}:
\begin{equation}\label{NVscaling}
\sigma^2(R) \sim \begin{cases}
R^{d-1}\ln R & \alpha = 1\\
R^{d-\alpha} & \alpha < 1\\
R^{d-1} & \alpha > 1
\end{cases}\qquad (R\rightarrow +\infty).
\end{equation}
Since disordered as well as ordered stealthy states can be viewed as systems
in which $\alpha$ tends to infinity, we see from Eq. (\ref{NVscaling}) that
they have the asymptotic scaling $\sigma^2(R) \sim R^{d-1}$. We give theoretical
predictions for the variance of disordered stealthy ground states in Sec. \ref{var}.

\section{Families of Stealthy Pair Potentials}
\label{family}

As we see in the next section,  the specific form of a stealthy potential does not affect the ground-state
energy manifold, but it can affect other thermodynamic properties, such as the
pressure. This has consequences in  simulations
of such properties, especially with respect to convergence issues.
Hence, it is instructive to remark on some mathematical aspects
of the long-range nature of the direct-space stealthy potentials, which
are very similar to the weakly decaying  Friedel oscillations of the electron density 
in a variety of systems, including molten metals  as well as graphene \cite{As76,Bac10}.
As we will see, in some cases, stealthy potentials may mimic effective interactions
that arise in certain  polymer systems \cite{Ha14}.

Here, we will limit ourselves to pair potentials $v(r)$ that are radial functions in $\mathbb{R}^d$,
where $r=|\bf r|$ (i.e., isotropic pair interactions), and therefore, their Fourier transforms ${\tilde v}(k)$ are also 
radial functions in  $\mathbb{R}^d$, where $k \equiv |{\bf k}|$ is a wave number. The $d$-dimensional Fourier transform
of any integrable radial function $f(r)$ in $\mathbb{R}^d$ is
given by \cite{To03a}
\begin{equation}
{\tilde f}(k) =\left(2\pi\right)^{\frac{d}{2}}\int_{0}^{\infty}r^{d-1}f(r)
\frac{J_{\left(d/2\right)-1}\!\left(kr\right)}{\left(kr\right)^{\left(d/2\right
)-1}}dr,
\label{FT}
\end{equation}
and the inverse transform of ${\tilde f}(k)$ is given by
\begin{equation}
f(r) =\frac{1}{\left(2\pi\right)^{\frac{d}{2}}}\int_{0}^{\infty}k^{d-1}{\tilde f}(k)
\frac{J_{\left(d/2\right)-1}\!\left(kr\right)}{\left(kr\right)^{\left(d/2\right
)-1}}dk,
\label{inverse}
\end{equation}
where $J_{\nu}(x)$ is the Bessel function of order $\nu$.

Consider the class of stealthy radial  potential functions  ${\tilde v}(k)$ in $\mathbb{R}^d$ 
that are bounded and  positive with compact support
in the radial interval $0 \le k \le K$, i.e., 
\begin{equation}
{\tilde v}(k)= V(k) \Theta(K-k),
\label{v-k}
\end{equation}
where, for simplicity, $V(k)$ is infinitely differentiable in the open interval $[0,K)$ and
\begin{equation}
\Theta(x) =\Bigg\{{0, \quad { x < 0}\atop{1, \quad x \ge 0}}
\label{step}
\end{equation}
is the Heaviside step function.  
The corresponding direct-space radial pair potential $v(r)$ is 
necessarily a delocalized, long-ranged function that is integrable in $\mathbb{R}^d$.
Moreover, without any loss of generality, it will be assumed that $V(k) \le v_0$.

For concreteness and purposes of illustration, we will examine properties of two specific families of potentials
that fall within the aforementioned wide class of  stealthy interactions: ``power-law" and ``overlap" potentials.

\subsection{Power-law potentials}

The power-law potentials are defined in Fourier space as follows:
\begin{equation}
{\tilde v}(k)=v_0 \,(1-k/K)^m\,\Theta(K-k),
\label{power}
\end{equation}
where the exponent $m$ can be any whole number.
The corresponding direct-space potential $v(r)$
will depend on $d$ for any given $m$ and is exactly given by
\begin{equation}
\frac{v(r)}{v_0}=\frac{K^d \,\Gamma(m+1)\Gamma((d+1)/2)\cdot \,_1F_2(a_1;b_1,b_2;x)}{\Gamma(1+m+d) \pi^{(d+1)/2} },
\label{direct-power}
\end{equation}
where $a_1=(d+1)/2$, $b_1=1+(m+d)/2$, $b_2=(1+m+d)/2$, $ x=-(Kr)^2/4 $
and $_1F_2(a_1;b_1,b_2;x)$ is a special case of the 
generalized hypergeometric function $_pF_q(a_1,...,a_p;b_1,...,b_q;x)$ \cite{Ab72}.
Because the potential (\ref{direct-power}) is derived from the Fourier power-law potential
(\ref{power}), we will  refer to Eq. (\ref{direct-power}) as the direct-space power-law potential.
In the instance when  $m=0$ in Eq. (\ref{power}) (i.e., simple
step function), this expression for $v(r)$ simplifies as follows:
\begin{equation}
\frac{v(r)}{v_0}= \left(\frac{K}{2 \pi r}\right)^{d/2}J_{d/2}(Kr),
\end{equation}
for which the large-$r$ asymptotic behavior is given by
\begin{equation}
\frac{v(r)}{v_0}\sim \left(\frac{K}{2 \pi }\right)^{(d-1)/2}
\frac{\cos(Kr -(d+1)\pi/4)}{\pi r^{(d+1)/2}}\qquad (r \rightarrow \infty).
\end{equation}
For any fixed $d$ and $m \ge d$, the direct-space power-law potential
has the asymptotic form
\begin{equation}
\frac{v(r)}{v_0}\sim \frac{s(r;m,d)}{r^{d+1}}\qquad (r \rightarrow \infty),
\end{equation}
where $s(r;m,d)$ is a bounded function (a sinusoidal function or constant
of order one). For any fixed $d$ and $1 \le m < d$, the long-range oscillations
of $v(r)$ are controlled by an envelope that decays like $1/r^{\beta}$, where
$(d+1)/2 < \beta \le d+1$.

\begin{figure}[H]
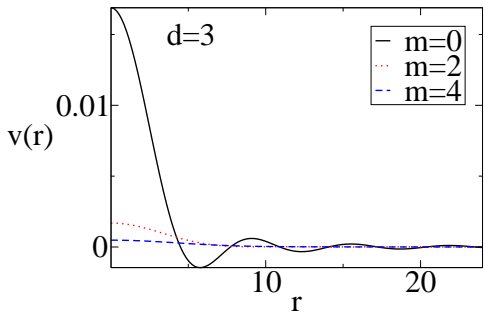

\begin{center}
{\includegraphics[  width=2.5in, keepaspectratio,clip=]{fig4a.eps} 
\includegraphics[  width=2.5in, keepaspectratio,clip=]{fig4b.eps}}
\caption{Left Panel: Fourier power-law potential ${\tilde v}(k)$ for the special cases
$m=0$, $2$, and $4$ that apply for any $d$. Right panel: Corresponding direct-space power-law potentials
$v(r)$ in the instance $d=3$. Here, we set $v_0=K=1$.}  \label{plot-power}
\end{center}
\end{figure}

In Fig. \ref{plot-power}, we plot the Fourier power-law potential
for selected values of $m$ (which applies in any dimension) and the corresponding direct-space potentials for $d=3$.
In all cases, we set $v_0=K=1$. It is to be noted that the amplitudes
of the oscillations in $v(r)$ decrease as $m$ increases for a fixed dimension.

\onecolumngrid

\twocolumngrid
\vspace{-0.4in}

\subsection{Overlap potentials}

Let  $\alpha(r;R)$ represent the intersection volume of two identical
$d$-dimensional spheres of radius $R$ (scaled by the volume of sphere) whose centers
are separated by a distance $r$. This quantity is known analytically
in any space dimension, and has a variety of representations \cite{To06b}, including the
following:
\begin{equation}
\alpha(r;R) = c(d) \int_0^{\cos^{-1}(r/(2R))} \sin^d(\theta) \, d\theta,
\label{alpha}
\end{equation}
where $c(d)$ is the $d$-dimensional constant given by
\begin{equation}
c(d)= \frac{2 \Gamma(1+d/2)}{\pi^{1/2} \Gamma((d+1)/2)}.
\end{equation}
For $d=1$, 2, 3 and 4, we respectively have
\begin{equation}
\alpha(r;R)=  \Theta(2R-r)\left[1 -\frac{r}{2R}\right]
\label{inter-v2-1d},
\end{equation}
\begin{equation}
\alpha(r;R)  =   \Theta(2R-r)\left[\frac{2}{\pi}
\left( \cos^{-1}\left(\frac{r}{2R}\right) - \frac{r}{2R} 
\left(1 - \frac{r^2}{4R^2}\right)^{1/2} \right)\right], 
\label{inter-v2-2d}
\end{equation}
\begin{equation}
\alpha(r;R)  =  \Theta(2R-r)\left[1 -\frac{3}{4} \frac{r}{R}+ \frac{1}{16}\left(\frac{r}{R}\right)^3\right] ,
 \label{inter-v2-3d} 
\end{equation}
\begin{eqnarray}
\alpha(r;R)&=& \Theta(2R-r) \nonumber \\
&&\hspace{-0.3in} \times\left[\frac{2}{\pi}
\left( \cos^{-1}\left(\frac{r}{2R}\right) - \left\{\frac{5r}{6R}-
\frac{1}{12}\left(\frac{r}{R}\right)^3\right\}  (1 - \frac{r^2}{4R^2})^{1/2} \right)\right].
\end{eqnarray}

Consider the class of ``overlap" potentials, which for any dimension
is given by
\begin{equation}
{\tilde v}(k)=v_0 \, \alpha(r=k,R=K/2).
\label{over}
\end{equation}
Note that for $d=1$, the overlap potential is identical to the power-law potential
when $d=1$ and $m=1$.
The thermodynamics of the ground-state manifold of this potential
in the special case $d=2$ was numerically investigated in Refs. \cite{Ba09a} and \cite{Ba09b}. 
It follows from Eq. (\ref{over}) that the corresponding direct-space overlap potential is given by
\begin{equation}
\frac{v(r)}{v_0}= \frac{\Gamma(1+d/2)}{\pi^{d/2}} \frac{J^2_{d/2}(Kr/2)}{r^d},
\end{equation}
which is clearly  {\it non-negative} for all $r$. Its
large-$r$ asymptotic behavior is given by
\begin{equation}
\frac{v(r)}{v_0} \sim \frac{4\Gamma(1+d/2)}{\pi^{d/2+1}r^{d+1}} \cos^2\left(Kr/2-\frac{(d+1)\pi}{4}\right) \qquad (r \rightarrow \infty),
\end{equation}
revealing that the long-ranged decay of the direct-space overlap potential
has an envelope controlled by the inverse power law $1/r^{d+1}$. 

\begin{figure}[ht]
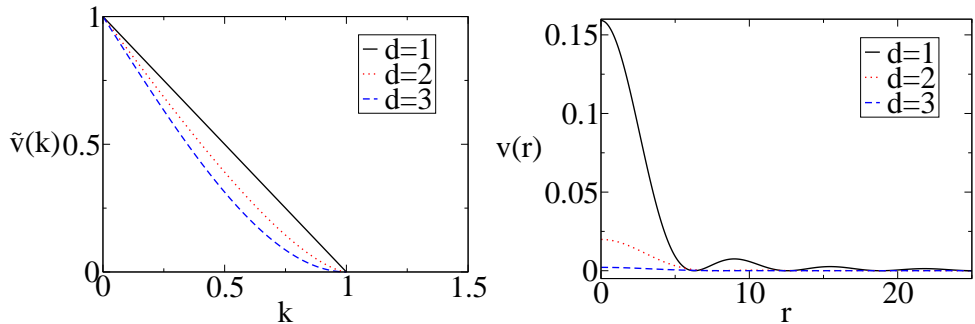

\begin{center}
{\includegraphics[  width=2.5in,clip=, keepaspectratio]{fig5a.eps} 
\includegraphics[  width=2.5in, clip=,keepaspectratio]{fig5b.eps}}
\caption{Left panel: Fourier overlap potential ${\tilde v}(k)$ for the first three
space dimensions. Right panel: Corresponding direct-space overlap potentials
$v(r)$, which oscillate but are always non-negative.  Here, we set $v_0=K=1$.}  
\label{plot-over}
\end{center}
\end{figure}

Figure \ref{plot-over}  depicts the overlap potential ${\tilde v}(k)$ for the first three
space dimensions and the corresponding direct-space overlap potentials
$v(r)$, the latter of which vividly shows the increasing decay rate of $v(r)$ with increasing 
dimension. The direct-space overlap potential $v(r)$ is similar in functional form to effective {\it positive} pair interactions
that arise in multilayered ionic microgels \cite{Ha14}.

\onecolumngrid

\twocolumngrid

\section{Ensemble Theory for Stealthy Disordered  Ground States: Exact Results for Thermodynamic Properties }
\label{ensemble}

Our general objective is the formulation of an ensemble theory for the 
thermodynamic and structural properties of stealthy degenerate disordered ground states that we previously investigated
numerically  \cite{Fa91,Uc04b,Uc06b,Ba08,Ba09a,Ba09b}. 
In this section, we derive general exact relations for thermodynamic properties
that apply to any well-defined ensemble as generated by a particular way
to sample the stealthy disordered ground-state manifold as a function of number density $\rho$.
In the subsequent section, we  derive some exact results for the pair statistics
for general ensembles.

\subsection{Preliminaries}

To begin, consider a configuration of $N$ identical particles with positions ${\bf r}^N \equiv 
{\bf r}_1,\ldots, {\bf r}_N$ in a large region of   volume $V$ 
in $d$-dimensional  Euclidean space $\mathbb{R}^d$. 
For particles interacting via a pair potential $v({\bf r})$, the 
total potential energy $\Phi({\bf r}^N)$ is given by
\begin{equation}
\Phi({\bf r}^N) =\sum_{i<j} v({\bf r}_{ij}),
\label{total}
\end{equation}
where ${\bf r}_{ij}={\bf r}_j-{\bf r}_i$.
Of particular interest are classical ground states, i.e., those configurations  
that minimize the energy per particle $\Phi({\bf r}^N)/N$.

The ensemble average of the energy (\ref{total}) per particle $u$ in the thermodynamic limit can be written in terms of  the pair correlation function
$g_2({\bf r})$:
\begin{eqnarray}
u \equiv \langle \frac{\Phi({\bf r}^N)}{N }\rangle
   &=&\frac{\rho}{2} \int_{\mathbb{R}^d}  v({\bf r})g_2({\bf r}) d{\bf r} \nonumber\\
&=&\frac{\rho}{2} \int_{\mathbb{R}^d} v({\bf r}) d{\bf r} +\frac{\rho}{2} \int_{\mathbb{R}^d} v({\bf r})h({\bf r}) d{\bf r},
\label{u1}
\end{eqnarray}
where angular brackets denote an ensemble average and $\rho$ is the number density in the thermodynamic limit. 
%i.e.,\begin{equation}
%\rho = \lim_{N,V \rightarrow \infty} \frac{N}{V}.
%\end{equation}

Because the collective-coordinate approach relies on the Fourier representation of the energy,
we recast Eq. (\ref{u1}) in terms of the structure factor $S({\bf k})$
by applying Parseval's theorem to the second line of Eq. (\ref{u1}):
\begin{eqnarray}
u&=&\frac{\rho}{2} {\tilde v}({\bf k=0})  +\frac{\rho}{2(2\pi)^d} \int_{\mathbb{R}^d} {\tilde v}({\bf k}){\tilde h}({\bf k}) d{\bf k} \nonumber\\
&=&\frac{\rho}{2} {\tilde v}({\bf k=0})  - \frac{1}{2} v({\bf r=0}) +\frac{1}{2(2\pi)^d} \int_{\mathbb{R}^d} 
{\tilde v}({\bf k})S({\bf k}) d{\bf k},
\label{u2}
\label{energy}
\end{eqnarray}
where ${\tilde v}({\bf k})$ and ${\tilde h}({\bf k})$ are the Fourier transforms of 
$v({\bf r})$ and $h({\bf r})$,  respectively, both of which are assumed to exist, $\bf k$ is a wave vector,
and $S({\bf k})$ is the structure factor defined in relation (\ref{factor}).
Note that the structure factor is a non-negative, inversion-symmetric function, i.e.,
\begin{equation}
S({\bf k}) \ge 0 \quad \mbox{for all} \; {\bf k}, \qquad S({\bf k}) = S(-{\bf k}).
\label{inv}
\end{equation}

\subsection{Ground-state energy and dimensionality of its configuration space}
\label{dim}

Consider a radial (isotropic) stealthy potential function ${\tilde v}(k)$ with support in $0 \le k \le K$
of the class specified by  Eq. (\ref{v-k}). In light of Eq. (\ref{u2}), it is clear that whenever 
particle configurations in $\mathbb{R}^d$ exist such that
$S(k)$ is constrained to achieve its minimum value of zero
for  $0 \le k \le K$, the system must be at its ground state or global energy minimum.
This follows because the integrand ${\tilde v}(k) S(k)$ in the nontrivial term on the right side of Eq. (\ref{u2})
is identically zero because of the conflicting demands of the step functions.  
 {When such configurations exist}, the average ground-state energy per particle
in any well-defined ensemble is given exactly by
\begin{equation}
u=\frac{\rho}{2} v_0 -\frac{1}{2}  v(r=0),
\label{ground}
\end{equation}
which is a {\it structure-independent} constant, depending on the density $\rho$,  $v_0 \equiv {\tilde v}(k=0)$,
$v(r=0)$ and $d$, as explicitly shown below. Importantly, because $u$ is a constant independent of
the structure, its value has no effect on the ground-state manifold, which is generally degenerate,
; hence, this manifold  is invariant to the specific choice of the stealthy function ${\tilde v}(k)$ at fixed $\rho$ and $d$.

We would like to express the energy (\ref{ground}) in terms of the parameter $\chi$, 
defined by relation (\ref{chi}), which measures
the relative number of independently constrained degrees of freedom
for a finite system under periodic boundary conditions. Note that, in the thermodynamic 
limit, $M(K)$ in  Eq. (\ref{chi}) is simply half of the volume of a sphere of radius $K$ [due to the inversion
symmetry of $S({\bf k})$] multiplied
by the  density  $\rho_{\Lambda^*}$ of the dual lattice [cf. Eq. (\ref{rho*})], i.e.,
\begin{equation}
M(K)= \rho_{\Lambda^*} \,\frac{v_1(K)}{2}=\frac{v_1(K)}{ 2(2\pi)^d \rho},
\end{equation}
where we have used the fact that $\rho \rightarrow N \rho_{\Lambda}$ in this
distinguished limit.  
Hence,  from Eq. (\ref{chi}), we obtain the following expression for $\chi$
in the thermodynamic limit:
\begin{equation}
\rho \,\chi =\frac{v_1(K)}{2d\,(2\pi)^d},
\label{rho-chi}
\end{equation}
where 
\begin{equation}
v_1(R)= \frac{\pi^{d/2} R^d}{\Gamma(1+d/2)}
\label{vol}
\end{equation}
is the volume of a $d$-dimensional sphere (hypersphere) of radius $R$.
We see that for fixed $K$ and $d$, which fixes the potential, $\chi$ is inversely proportional to $\rho$,
which is the situation that we usually consider in this paper \cite{Note6}.

Hence, as $\chi$ tends to zero, $\rho$ tends to infinity, which configurationally corresponds counterintuitively to the
uncorrelated ideal-gas limit (Poisson distribution), as discussed in the Introduction.
As $\chi$ increases from zero, the density $\rho$ decreases and the dimensionality
of the ground-state configuration manifold $D_C$ decreases. The configurational dimensionality per particle
in the thermodynamic limit, $d_C$, can easily be obtained from the relation $D_C=dN -2 M(K)$
for a finite system \cite{Note7}; specifically,
\begin{equation}
d_C =d(1-2\chi),
\label{dc}
\end{equation}
where $d_C = \lim_{D_C \rightarrow \infty,N \rightarrow \infty} D_C/N$.

 {Equations  (\ref{ground}) and (\ref{rho-chi}})  yield  the average ground-state energy per particle to be
\begin{equation}
u=v_0\left[\frac{\rho}{2}  - \gamma d\rho \chi\right], \qquad (\rho^*_{min} \le \rho < \infty),
\label{ground2}
\end{equation}
where $\rho_{min}^*$ is the minimal density associated with
the dual of the densest Bravais lattice in direct space  (as elaborated in Sec. \ref{exist}), and
\begin{equation}
\gamma= \frac{\int_{\mathbb{R}^d} {\tilde v}(k) d{\bf k}}{v_0 v_1(K)}=\frac{(2\pi)^d v(r=0)}{v_0 v_1(K)}
\label{gamma}
\end{equation}
is a constant whose value depends on the specific form of the stealthy-potential class ${\tilde v}(k)$
defined by Eq. (\ref{v-k}) and hence  must lie
in the interval $(0,1]$, where $\gamma=1$ corresponds to the step-function choice ${\tilde v}(k)=v_0 \Theta(K-k)$.
%However, to reiterate, because this is a structure-independent constant, it 
%has no effect on the ground-state manifold.
While the system in the limit $\chi \rightarrow 0$ ($\rho \rightarrow \infty$) corresponds configurationally
to an ideal gas  {in so far as the pair correlation function is concerned, as we will explain in detail in Sec.~\ref{subsec7a}, } thermodynamically, it is nonideal; see Eq. (\ref{ground2}) for $u$
and Eq. (\ref{p-u}) for the pressure.

\subsection{Energy route to pressure and isothermal compressibility}

The pressure in the thermodynamic limit at $T=0$ can be obtained from the
energy per particle via the relation
\begin{equation}
p= \rho^2 \left(\frac{\partial u}{\partial \rho}\right)_T.
\end{equation}
Therefore, for stealthy potentials, we see from Eq. (\ref{ground}) that
the ground-state pressure, for all possible values of $\rho$ or $\chi$, is given by the
following simple expression:
\begin{equation}
p= \frac{\rho^2}{2}v_0, \qquad (\rho^*_{min} \le \rho < \infty).
\label{p-u}
\end{equation}
Hence, the isothermal compressibility $\kappa_T\equiv \rho^{-1}\left(\frac{\partial \rho}{\partial p}\right)_T$
of such a ground state is
\begin{equation}
\kappa_T=\frac{v_0}{\rho^2}.
\label{iso}
\end{equation}
We see that as $\rho$ tends to infinity, the compressibility tends to zero.

Two important remarks are in order. First, estimates of the pressure
obtained from simulations that we previously performed for $d=2$
\cite{Ba09a,Ba09b}, as well as those carried
out in the present study across the first three space dimensions (Appendix B),
are in very good agreement  with the exact result  (\ref{p-u}) across a wide range of densities,
thus validating the accuracy of the simulations. Second, the fact that the pressure (\ref{p-u})
is a continuous function of density implies that any phase transition that may take place
 {could be } a continuous one, the implications of which are  discussed in the Conclusions.

\subsection{Virial route to pressure and isothermal compressibility}

An alternative route to the pressure for a radial pair potential function $v(r)$ is through the ``virial" equation, which 
at $T=0$ in the thermodynamic limit, is given by
\begin{equation}
p= -\frac{\rho^2}{2d} s_1(1) \int_0^\infty r^d \; \frac{d v}{d r} g_2(r) dr.
\label{virial1}
\end{equation}
Although the pressure obtained via the virial route is generally expected to be equivalent to that obtained from the energy route (as described in the previous section), we will show that, for a certain class of stealthy potentials, the pressure obtained  from Eq. \eqref{virial1} is either ill defined or divergent. This has practical implications for what types of stealthy potentials can be used in constant-pressure simulations. 
 
 It is convenient to rewrite the virial relation (\ref{virial1}) in the following form:
\begin{eqnarray}
p&=& -\frac{\rho^2}{2d}\left[ \int_{\mathbb{R}^d}  r\frac{d v}{d r} d{\bf r} + \int_{\mathbb{R}^d}  r\frac{d v}{d r} \;h(r) d{\bf r}\right] \nonumber\\ 
&=& -\frac{\rho^2}{2d}\left[ {\tilde F}(k=0) + \frac{1}{(2\pi)^d}\int_{\mathbb{R}^d} {\tilde F}(k)  \;{\tilde h}(k) d{\bf k}\right]
\label{virial2}
\end{eqnarray}
where ${\tilde F}(k)$ is the Fourier transform of $F(r) \equiv r dv/dr$, when it exists, and we have used Parseval's theorem and  definition (\ref{tot}) for the total correlation function $h(r)$. 

To continue with this analysis, we make use of the following lemma.

Consider a bounded radial function ${\tilde z}(k)$ with compact support on the radial interval  $[0,K]$ in $\mathbb{R}^d$ that
is infinitely differentiable in the open interval $[0,K)$. Therefore, its Fourier transform $z(r)$ exists.

{\sl Lemma 1.---}The Fourier transform of the radial function $w(r)=r dz/dr$ in $\mathbb{R}^d$
is given by
\begin{equation}
{\tilde w}(k)= - d \cdot {\tilde z}(k) - \frac{d {\tilde z}}{d k}.
\end{equation}

\noindent Proof.---Differentiation of ${\tilde z}(k)$ [defined via Eq. (\ref{FT})] with respect to $k$ leads
to the following identity:
\begin{equation}
k \frac{d {\tilde z}}{dk}=-\left(2\pi\right)^{\frac{d}{2}}\int_{0}^{\infty}k r^{d}z(r)
\frac{J_{\left(d/2\right)}\!\left(kr\right)}{\left(kr\right)^{\left(d/2\right
)-1}}dr.
\label{lemma1}
\end{equation}
The Fourier transform of $w(r)$ is given by
\begin{equation}
{\tilde w}(k)=\left(2\pi\right)^{\frac{d}{2}}\int_{0}^{\infty}r^{d}
\frac{d z}{d r}
\frac{J_{\left(d/2\right)-1}\!\left(kr\right)}{\left(kr\right)^{\left(d/2\right
)-1}}dr.
\label{lemma2}
\end{equation}
Integrating relation (\ref{lemma2}) by parts and using Eq. (\ref{lemma1}) proves the lemma.

{\sl Corollary.---}It immediately follows from Lemma 1 that ${\tilde w}(k)$ has the same support as ${\tilde z}(k)$ and 
\begin{equation}
{\tilde w}(k=0)= - d \cdot {\tilde z}(k=0),
\end{equation}
meaning that the volume integral of $r dz/dr$ over all space is proportional to the corresponding 
volume integral of $z(r)$.

Note that by the Corollary of Lemma  1, ${\tilde F}(0)=-d\cdot {\tilde v}(0)$, and hence
we can rewrite the virial relation (\ref{virial2}) as
\begin{eqnarray}
p&=& \frac{\rho^2}{2d}\left[ d {\tilde v}(k=0) + \frac{d}{(2\pi)^d}\int_{\mathbb{R}^d} {\tilde F}(k)  \;{\tilde h}(k) d{\bf k}\right]\nonumber \\
&=& \frac{\rho^2}{2d}\left[ d {\tilde v}(k=0) - \frac{d}{\rho(2\pi)^d}\int_{\mathbb{R}^d} {\tilde F}(k)  d{\bf k}\right]\nonumber \\
&=& \frac{\rho^2}{2} v_0.
\label{virial3}
\end{eqnarray}
The second term in the second line of Eq. (\ref{virial3}) follows because
$\rho{\tilde h}(k)=-\Theta(K-k)$  inside the exclusion sphere of radius $K$ [see also Eq. (\ref{h-FT}) below]
and has support in this exclusion zone by the Corollary of Lemma  1. But this second term
must vanish in light of the trivial identity
\begin{equation}
F(r=0)= \left(r \frac{d v}{d r}\right)_{r=0}=\frac{1}{(2\pi)^d} \int_{\mathbb{R}^d} {\tilde F}(k) d{\bf k}=0.
\label{1}
\end{equation}
We see that the virial ground-state pressure (\ref{virial3}) agrees with that of the pressure obtained 
via the energy per particle [cf. Eq. (\ref{p-u})] {\it provided} that ${\tilde F}(k)$ exists. Since the latter is a stronger
condition than the existence of ${\tilde v}(k)$, it is possible to devise a stealthy function ${\tilde v}(k)$ for which
${\tilde F}(k)$ does not exist and hence a virial pressure that either diverges or is nonconvergent.
For example, this problem occurs for the power-law potential (\ref{power}) with $m=0$ (step function) for
any dimension $d$. By contrast, the virial pressure is always well defined for
the overlap potential (\ref{over}) in any dimension \cite{Note8}. This example serves to illustrate
the mathematical subtleties that can arise because of the long-ranged nature
of stealthy potentials in direct space.

\section{Ensemble Theory for Stealthy Disordered  Ground States: Exact Integral Conditions on the Pair Statistics}
\label{pair}

Here, we derive some exact integral conditions that must be obeyed by both the pair correlation function $g_2(r)$
and the structure factor $S(k)$ for stealthy ground states that apply to general
ensembles. These analytical relations can be profitably employed to test corresponding
computer simulation results.

\subsection{General properties}
\label{pair-1}

In any stealthy ground state, the structure factor
attains its minimum value $S(k)=0$ for $0 < k \le K$, and hence has the form
\begin{equation}
S(k)=\Theta(k-K)[1+{\tilde Q}(k)],
\end{equation}
where $\Theta(x)$ is the Heaviside step function defined by Eq. (\ref{step})
and ${\tilde Q}(k)=S(k)-1$ is a function that obeys the inequality $ {\tilde Q}(k)\ge -1$.
Therefore, from Eq. (\ref{factor}), we have that the Fourier transform of the total correlation function $h(r)$ has the form 
\begin{equation}
\rho {\tilde h}(k) ={\tilde f}(k) + {\tilde P}(k),
\label{h-FT}
\end{equation}
where
\begin{equation}
{\tilde f}(k)=-\Theta(K-k)  
\end{equation}
and
\begin{equation}
{\tilde P}(k)=\Theta(k-K)  {\tilde Q}(k).
\end{equation}
It is noteworthy that the function ${\tilde f}(k)$ is identical to the Mayer-$f$ function
for an equilibrium hard-sphere system in direct space.

Taking the inverse Fourier transform of Eq. (\ref{h-FT}) yields the direct-space
total correlation function, given by
\begin{equation}
\rho h(r)=f(r)+ P(r)
\end{equation}
where 
\begin{equation}
f(r)= -\left(\frac{K}{2\pi  r}\right)^{d/2} J_{d/2}(Kr) 
\end{equation}
and
\begin{eqnarray}
P(r)&=&\frac{1}{\left(2\pi\right)^{\frac{d}{2}}}\int_{K}^{\infty}k^{d-1}{\tilde Q}(k)
\frac{J_{\left(d/2\right)-1 }\!\left(kr\right)}{\left(kr\right)^{\left(d/2\right
)-1}}dk \nonumber \\
&\ge& \left(\frac{K}{2\pi  r}\right)^{d/2}J_{d/2}(Kr)   -\rho,
\label{P}
\end{eqnarray}
where the lower bound on $P(r)$ indicated in Eq. (\ref{P}) follows from 
the fact that $h(r) \ge -1$ for all $r$ for any point pattern.
It trivially follows that since $\rho {\tilde h}(k=0)=\rho \int_{\mathbb{R}^d} h(r) d{\bf r}=-1$,
the volume integral of $P(r)$ must be zero, i.e.,
\begin{equation}
\int_{\mathbb{R}^d} P(r) d{\bf r} =0 .
\label{cond1}
\end{equation}
Less trivially, because the product ${\tilde v}(k) {\tilde P}(k)$ is zero for all $k$, 
by Parseval's theorem, we have the integral  condition 
\begin{equation}
\int_{\mathbb{R}^d} v(r) P(r) d{\bf r} =0 .
\label{cond2}
\end{equation}
Thus, the functions $v(r)$ and $P(r)$ are orthogonal to one another.
The exact integral conditions (\ref{cond1}) and (\ref{cond2})
can be used to test the accuracy of numerical methods
that yield estimates of the pair correlation function.

\subsection{Behavior of the pair correlation function near the origin}
\label{pair-2}

It is instructive to determine the behavior of the pair correlation function $g_2(r)$
for small $r$. Substitution of the general form (\ref{FT}) for ${\tilde h}(k)$ into the definition
of the total correlation function $h(r)$ as obtained from Eq. (\ref{inverse}), and expanding $h(r)$ in
a Taylor series around $r=0$ through second order in the radial distance $r$, yields
\begin{equation}
h(r) = h(r=0) + \frac{1}{2}\left(\frac{\partial^2 h}{\partial r^2}\right)_{r=0}  r^2 +{\cal O}\left(r^4\right),
\end{equation}
where 
\begin{equation}
h(r=0)=-2d \chi + 2d^2\chi \int_{K}^{\infty} k^{d-1} {\tilde Q}(k) dk
\label{origin}
\end{equation}
and the corresponding curvature is
\begin{equation}
\left(\frac{\partial^2 h}{\partial r^2}\right)_{r=0}=
\frac{2d}{d+2} \chi  -2d \chi \int_{K}^{\infty} k^{d+1} {\tilde Q}(k) dk.
\label{curve}
\end{equation}

Therefore, from Eq. (\ref{origin}), we see that the pair correlation function at the origin is given by
\begin{equation}
g_2(r=0)=1-2d \chi +2d^2\chi \int_{K}^{\infty} k^{d-1} {\tilde Q}(k) dk
\end{equation}
Since $g_2(r)$ must be non-negative for all $r$, we have the following integral condition
on ${\tilde Q}(k)$:
\begin{equation}
2d^2\chi\int_{K}^\infty k^{d-1} {\tilde Q}(k) dk \ge 2d\chi -1.
\label{ineq1}
\end{equation}
Hence, this integral must be positive for
\begin{equation}
\chi \ge \frac{1}{2d}.
\end{equation}
We also conclude from Eq. (\ref{curve}) that for $h(r)$ or $g_2(r)$ to have positive curvature
at the origin, ${\tilde Q}(k)$ must obey the additional integral condition:
\begin{equation}
(d+2) \int_{K}^{\infty} k^{d+1} {\tilde Q}(k) dk \le 1.
\label{ineq2}
\end{equation}
Finally, we note that when $g_2(r=0)=0$, the results above yield
the equality
\begin{equation}
\int_{K}^\infty k^{d-1} {\tilde Q}(k) dk = \frac{2d\chi -1}{2d^2\chi}
\label{eq}
\end{equation}
and, because the curvature must be positive in this instance, the inequality
(\ref{ineq2}) must generally be obeyed. 

The inequality (\ref{ineq1}), conditional inequality (\ref{ineq2}),
and conditional equality (\ref{eq}) provide 
integral conditions to test the accuracy of numerical methods
that yield estimates of the structure factor.

\section{Existence of Stealthy Disordered Degenerate Ground States}
\label{exist}

It is noteworthy that any periodic crystal with a finite basis is a  stealthy ground state
for all positive $\chi$ up to its corresponding  maximum value $\chi_{max}$ (or minimum
value of the number density $\rho_{min}$) determined by its first positive Bragg peak ${\bf k}_{Bragg}$
[minimal positive wave vector for which $S({\bf k})$ is positive].
Tables I-IV list the pair $\chi_{max}$,$\rho_{min}$ for some common 
periodic patterns in one, two, three, and four dimensions, respectively, all
of which are part of the ground-state manifold; see Appendix A
for mathematical definitions. (The crystals denoted by $\mbox{Dia}_d$ and $\mbox{Kag}_d$
are $d$-dimensional generalizations of the diamond and  kagom{\' e} crystals, respectively, for $d\ge 2$  \cite{Za11e}.)
While the mere existence of such periodic ground states does not provide any clues
about their occurrence probability in some ensemble, we will use these results  here
to show how disordered degenerate ground states arise as part of the ground-state manifold
for sufficiently small $\chi$.

\begin{table}[H]
\caption{\label{1D}Maximum values of $\chi$  
and  corresponding minimum values of $\rho$ for certain
periodic  stealthy ground states in $\mathbb{R}$ with $K=1$. The configuration
with the largest possible value of $\chi_{max}$ (smallest possible value
of $\rho_{min}$) corresponds to the integer lattice.}
\begin{center}
\begin{tabular}{c|c|c}
Structure                               & $\chi_{max}$ & $\rho_{min}$ \\ \hline
Integer lattice    ($\mathbb{Z}$)     & $1$ & $\frac{1}{2\pi}=0.15915\ldots$   \\
Periodic with $n$-particle basis    & $\frac{\displaystyle 1}{\displaystyle n}$ & $\frac{n}{2\pi}=(0.15915\ldots)n$        \\
\end{tabular}
\end{center}
\end{table}

\onecolumngrid

\begin{table}[H]
\caption{\label{2D}Maximum values of $\chi$  
and  corresponding minimum values of $\rho$ for certain
periodic  stealthy ground states in $\mathbb{R}^2$ with $K=1$. The configuration
with the largest possible value of $\chi_{max}$ (smallest possible value
of $\rho_{min}$) corresponds to the triangular lattice.}
\begin{center}
\begin{tabular}{c|c|c}
Structure                               & $\chi_{max}$ & $\rho_{min}$ \\ \hline
Kagom\'{e}  crystal ($\mbox{Kag}_2$)         & $\frac{\pi}{3\sqrt{12}}=0.3022\ldots$ & $\frac{3\sqrt{3}}{8\pi^2}=0.06581\ldots$   \\
Honeycomb  crystal  ($\mbox{Dia}_2$)        & $\frac{\pi}{2\sqrt{12}}=0.4534\ldots$ & $\frac{\sqrt{3}}{4\pi^2}=0.04387\ldots$  \\
Square lattice    ($\mathbb{Z}^2=\mathbb{Z}^2_*$)             & $\frac{\pi}{4}= 0.7853\ldots$ &  $\frac{1}{4\pi^2}=0.02533\ldots$ \\
Triangular lattice     ($A_2\equiv A_2^*$)             & $\frac{\pi}{\sqrt{12}}=0.9068\ldots$ & $\frac{\sqrt{3}}{8\pi^2}=0.02193\ldots$\\
\end{tabular}
\end{center}
\end{table}

\begin{table}[H]
\caption{\label{3D} Maximum values of $\chi$  
and  corresponding minimum values of $\rho$ for certain periodic  stealthy ground states  in $\mathbb{R}^3$ with $K=1$.
Here MCC refers to the mean-centered cuboidal lattice, which is a Bravais lattice intermediate
between the BCC and FCC lattices, and  has an equivalent dual lattice \cite{Co94}. The configuration
with the largest possible value of $\chi_{max}$ (smallest possible value
of $\rho_{min}$) corresponds to the BCC lattice.}
\begin{center}
\begin{tabular}{c|c|c}
Structure                               & $\chi_{max}$ & $\rho_{min}$ \\ \hline
Pyrochlore crystal ($\mbox{Kag}_3$)         & $\frac{\pi}{4\sqrt{12}}=0.2267\ldots$ & $\frac{2}{3\sqrt{3}\pi^3}=0.01241\ldots$    \\
Diamond  crystal ($\mbox{Dia}_3$)       & $\frac{\pi}{2\sqrt{12}}=0.4534\ldots$ & $\frac{1}{3\sqrt{3}\pi^3}=0.00620\ldots$    \\
Simple hexagonal lattice  & $\frac{\sqrt{3}\,\pi}{9}= 0.6045\ldots$ & $\frac{1}{4\sqrt{3}\,\pi^3}=0.00465\ldots$ \\
SC lattice  ($Z_3 \equiv Z_3^*$)             & $\frac{2\pi}{9}= 0.6981\ldots$ & $\frac{1}{8\pi^3}=0.00403\ldots$  \\
HCP crystal               & $\frac{8\sqrt{6}\pi}{81}= 0.7600\ldots$ & $\frac{3\sqrt{3}}{32\sqrt{2}\pi^3}=0.00370\ldots$ \\
FCC lattice ($D_3\equiv A_3$)            & $\frac{\pi}{\sqrt{12}}= 0.9068\ldots$ &  $\frac{1}{6\sqrt{3}\pi^3}=0.00310\ldots$ \\
MCC lattice         & $0.9258\ldots$ & $0.00303\ldots$    \\
BCC lattice ($D_3^* \equiv D_3^*$)           & $\frac{2\sqrt{2}\pi}{9}=0.9873\ldots$ &  $\frac{1}{8\sqrt{2}\pi^3}=0.00285\ldots$ \\
\end{tabular}
\end{center}
\end{table}
\twocolumngrid

\begin{table}[H]
\caption{\label{4D} Maximum values of $\chi$  
and  corresponding minimum values of $\rho$ for certain periodic  stealthy ground states  in $\mathbb{R}^4$ with $K=1$.
The configuration
with the largest possible value of $\chi_{max}$ (smallest possible value
of $\rho_{min}$) corresponds to the four-dimensional checkerboard lattice $D_4 \equiv D_4^*$.}
\begin{center}
\begin{tabular}{c|c|c}
Structure                               & $\chi_{max}$ & $\rho_{min}$ \\ \hline
$\mbox{Kag}_4$ crystal          & $\frac{\pi^2}{40}=0.2467\ldots$& $\frac{5}{32\pi^3}=0.001640\ldots$   \\
$\mbox{Dia}_4$ crystal         & $\frac{\pi^2}{16}=0.6168\ldots$& $\frac{1}{16\pi^3}=0.0006416\ldots$   \\
$\mathbb{Z}^4$ lattice         & $\frac{\pi^2}{16}=0.6168\ldots$& $\frac{1}{16\pi^3}=0.0006416\ldots$   \\
$D_4$ lattice         & $\frac{\pi^2}{8}=1.2337\ldots$& $\frac{1}{32\pi^3}
=0.0003208\ldots$   \\

\end{tabular}
\end{center}
\end{table}

At fixed $d$, the smallest value of $\rho_{min}$ listed in Tables I-IV, which we call $\rho_{min}^*$, corresponds
to the dual of the densest Bravais lattice in direct space, and 
represents the critical density value below which a stealthy ground state
does not exist for all $k \le |{\bf k}_{Bragg}^*|$.  The fact that $\rho_{min}^*$ corresponds to the body-centered-cubic 
(BCC) lattice for $d=3$ was initially shown analytically in  Ref. \cite{Su05} 
and subsequently numerically in Ref. \cite{Ba08}. We note that the values
of $\rho_{min}$ for the simple hexagonal lattice and hexagonal
close-packed crystal for $d=3$ reported in Ref.  \cite{Su05} are incorrect
because those calculations were based on the erroneous assumption
that the structure factors at the corresponding shortest reciprocal lattice vectors have  nonvanishing 
values.

Observe that in the case $d=1$, there is no non-Bravais lattice (periodic structure
with a basis $n \ge 2$) for which $\chi_{max}$ is greater than $1/2$, implying
that the ground-state manifold is nondegenerate (uniquely the integer lattice) for $1/2 < \chi \le 1$.
This case is to be contrasted with the cases $d \ge 2$ where the ground-state manifold must be degenerate \cite{Note9}
for $1/2 <\chi < \chi_{max}^*$ and nondegenerate only at the point $\chi=\chi_{max}^*$,
as implied by Tables II-IV. Here, $\chi_{max}^*$ is the largest possible value of $\chi_{max}$
in some fixed dimension.\smallskip

\noindent{\sl Lemma.---}At fixed $K$, a configuration comprised of the union (superposition) of $m$ different stealthy ground-state
configurations in $\mathbb{R}^d$ with $\chi_1,\chi_2,\ldots,\chi_m$, respectively,
is itself stealthy with a $\chi$ value given by
\begin{equation}
\chi= \left[\sum_{i=1}^m \chi_i^{-1}\right]^{-1},
\label{harm}
\end{equation}
which is the harmonic mean of the $\chi_i$ divided by $m$.

\indent{\it Proof.---}Formula (\ref{harm}) is a direct consequence of the fact that $\chi$ is inversely
proportional to the number density $\rho=\sum_{i=1}^m \rho_i$ of the union of the configurations in $\mathbb{R}^d$,
where $\rho_i$  is the number density associated with the $i$th configuration, which
is inversely proportional to $\chi_i$.

%In the case that $\chi_i=\chi_0$ for all $i$, we have the simpler result
%\begin{equation}
%\chi= \frac{\chi_0}{m}.
%\end{equation}

This Lemma, together with the fact that any periodic crystal with a finite basis
is a stealthy ground state  can be used to demonstrate rigorously how complex 
aperiodic patterns can be ground states, entropically favored or not. A sketch of such a proof
would involve the   consideration of the union
of $m$ different periodic structures in $\mathbb{R}^d$ with densities $\rho_1,\rho_2,\ldots,\rho_m$, respectively,
each of which are randomly translated and oriented with respect to some coordinate system
such that $m$ is very large but bounded and $\rho_i \neq \rho_j$ for all $i$ and $j$.  
It is clear that the resulting configuration will be
a highly complex {\it aperiodic} structure in $\mathbb{R}^d$ that tends toward a disordered 
stealthy pattern with a value of $\chi$ that is very small but positive according
to relation (\ref{harm}).

\section{Pair Statistics in the Canonical Ensemble: ``Pseudo" Hard Spheres in Fourier Space}
\label{pseudo}

The task of formulating an ensemble theory that yields
analytical expressions for the pair statistics of stealthy degenerate
ground states is highly nontrivial because the dimensionality of the  configuration
space depends on the density (or $\chi$) and there is a multitude of ways
of sampling the ground-state manifold, each with its own probability measure
for finding a particular ground-state configuration. Therefore, 
it is desirable to specialize to  equilibrium ensembles with Gibbs measures because
the characterization of the ground states (as well as the corresponding excited states) 
would be  most tractable theoretically.  In particular,
our objective is to derive analytical formulas for the pair statistics
of stealthy disordered ground states for sufficiently small $\chi$ 
in the canonical ensemble as temperature $T$ tends to zero; i.e.,
the probability of observing a configuration is proportional
to $\exp[-\Phi({\bf r}^N)/(k_B T)]$ in the limit $T \rightarrow 0$.
We show here that under such circumstances, the pair statistics in the thermodynamic limit
can be derived under the ansatz that stealthy ground states behave remarkably like pseudo-equilibrium  hard-sphere systems in Fourier space.
This ansatz enables us to exploit well-known accurate expressions
for the pair statistics in direct space. As will be shown, agreement with computer simulations
is excellent for sufficiently small $\chi$.

\subsection{``Pseudo" hard-sphere ansatz}
 \label{subsec7a}

We have already noted that the step-function contribution to $\rho {\tilde h}(k)$ for  stealthy
ground states, denoted by ${\tilde f}(k)$ in relation (\ref{h-FT}), is identical to the Mayer-$f$ function
for an equilibrium hard-sphere system in direct space. This implies
that the corresponding contribution to $S(k)$ is a simple hard-core step function $\Theta(k-K)$,
which can be viewed as an equilibrium  hard-sphere system in Fourier space
with ``spheres" of diameter $K$ in the limit that $\chi$ tends to zero.
Why is this the case? Because such a step function is exactly the same as the pair correlation $g_2(r=k)$
of an equilibrium hard-sphere system in direct space in the limit that $\rho$ tends to zero. 
%The structure factor of  a stealthy ground state in the limit
%$\chi \rightarrow 0$ for $S(k)$ is the same \textcolor{blue}{dilute limit $\rho \rightarrow 0$} 
%of $g_2(r)$ for a hard-sphere system in direct space. 
That the structure factor must have the behavior 
$S(k) \rightarrow \Theta(k-K)$ in the limit $\chi \rightarrow 0$ is perfectly reasonable, since
a perturbation about the ideal-gas limit [where $S(k)=1$ for all $k$] in which an infinitesimal
fraction of the degrees of freedom are constrained should only introduce an infinitesimal change in $S(k)$
of zero inside the exclusion zone (constrained region). We call this the weakly constrained
limit, where a step function $S(k)$ is expected on maximum entropy grounds;
it corresponds to the most disordered (decorrelated) form of $S(k)$ subject to the impenetrability condition
in Fourier space. We refer to this phenomenon as equilibrated pseudo hard spheres in Fourier space because
there are actually no points in that space that have a hard-core repulsion like true hard spheres
do in direct space.

%but yet $S(k)$ for $\chi \rightarrow 0$ is functionally identical to an equilibrium hard-sphere $g_2$ 
%in direct space in the limit $\rho \rightarrow 0$. 

On the same maximum entropy grounds, we expect that a perturbation expansion about the
weakly constrained limit $\chi=0$ will lead to a perturbation expansion in $\chi$ for $S(k)$
that can be mapped to the low-density expansion of $g_2(r)$ for equilibrium hard spheres. 
More generally, we make the ansatz that, in the canonical ensemble as $T \rightarrow 0$, 
this hard-sphere analogy continues to hold as $\chi$ is increased
from zero to positive values, provided that $\chi$ is small enough, implying that
the collective coordinate variables ${\tilde n}(k)$ (defined in the Introduction)  are weakly correlated.
Though the pseudo-hard-sphere picture must break down in some intermediate range of $\chi$, for $d=1$
and $d=2$, this hard-sphere mapping is again exact when $\chi=\chi_{max}$, which
corresponds to the maximal value of the packing fraction $\eta$ in these dimensions
(see Tables I and II). This exact correspondence with the maximal value of $\eta$ when  $\chi=\chi_{max}$
does not hold for $d=3$ or $d=4$, however. Thus, one should only expect that $\chi$
and $\eta$ are proportional to one another, even at small $\chi$ values.

Under the pseudo-hard-sphere ansatz, 
%we now consider carrying out a series expansion of the structure factor $S(k)$ about $\chi=0$.
%Motivated by the fact that $S(k)$ is a  step-function in the limit $\chi \rightarrow 0$,
%we make the ansatz that for sufficiently small $\chi$, 
the direct-space pair correlation function $g_2^{HS}(r;\eta)$ of a disordered hard-sphere system 
at a packing fraction $\eta$ for sufficiently small $\eta$ can be mapped
into the structure factor $S(k;\chi)$ for a disordered stealthy ground state derived from the canonical
ensemble at fixed $\chi$ for sufficiently small $\chi$ as follows:
\begin{equation}
S(k;\chi)=g_2^{HS}(r=k;\eta).
\label{eq_sk}
\end{equation}
As alluded to above, the parameter $\chi$ can be viewed as an 
effective packing fraction for pseudo hard spheres of diameter $K$ in reciprocal space
that is proportional to $\eta$, i.e.,
\begin{equation}
\eta=b(d)\chi,
\label{b}
\end{equation}
where $b(d)$ is a $d$-dependent parameter that  is to be determined.
Let $h_{HS}(r)$ be the total correlation function of a disordered equilibrium hard-sphere 
system in direct space and  let us define for stealthy ground states
\begin{equation}
{\tilde H}(k) \equiv S(k)-1=\rho {\tilde h}(k).
\label{H1}
\end{equation}
The ansatz is also defined by the alternative mapping
\begin{equation}
{\tilde H}(k)=h_{HS}(r=k).
\label{psuedo}
\end{equation} 
This mapping then enables us to exploit
the well-known statistical-mechanical theory of equilibrium hard-sphere systems.
In particular, we can employ a  generalized Ornstein-Zernike convolution relation that defines
the appropriate direct correlation function ${\tilde C}(k)$, namely, 
\begin{equation}
{\tilde H}(k) = {\tilde C}(k) + \eta\, {\tilde H}(k) \otimes {\tilde C}(k),
\label{OZ2}
\end{equation}
where  the symbol $\otimes$ denotes the convolution operation in $\mathbb{R}^d$.
Therefore, in  direct space, $H(r)$ is given by the relation of the following form:
\begin{equation}
H(r) =\frac{C(r)}{1-(2\pi)^d \,\eta\, C(r)}.
\end{equation}

For example, for $d=1$,
\begin{equation}
{\tilde C}(k)=-\Theta(1-k) \frac{(1-\eta k)}{(1-\eta)^2}.
\end{equation}
Inverting this function yields
\begin{equation}
C(r)= \frac{- r\sin(r) + ( r [\sin(r) +\cos(r)] -1) \eta}{\pi r^2 (1-\eta)^2}
\end{equation} 
For $d=2$ and $d=3$, one can use
the Percus-Yevick closure of the Ornstein-Zernike integral equation \cite{Ha86},
which is highly accurate for low to intermediate densities along the liquid branch, or
when mapped to the stealthy problem, for low to intermediate values
of $\chi$.

It is noteworthy that the exact low-density expansion
of $h_{HS}(r)$, for practical purposes, is sufficient to produce accurate estimates
of ${\tilde H}(k)=\rho {\tilde h}(k)$ and its counterpart $\rho h(r)$ for low to intermediate values of $\chi$ or $\eta$. In 
particular, using the mapping (\ref{psuedo}), we obtain,
for any dimension $d$, the following low-$\chi$ expansion
of $ \rho{\tilde h}(k)$ :
\begin{equation}
\rho {\tilde h}(k) = -\Theta(K-k) \left[ 1+ 2^d \, b(d) \, \alpha(k;K) \, \chi  + {\cal O}(\chi^2)\right],
\label{H2}
\end{equation}
where $b(d)$ is the proportionality constant in Eq. (\ref{b}) and $\alpha(k;K)$ is the scaled intersection volume of two identical
$d$-dimensional spheres of diameter $K$ whose centers
are separated by a distance $k$ [cf. Eq. (\ref{alpha})] \cite{Note10}.
This formula indicates that $S(k)$ develops a peak value at $k=K$ (over and above the value of unity
due to the step function in the limit $\chi \rightarrow 0$) and then monotonically decreases
until $k=2K$, where it achieves its long-range value of unity for all $k> 2K$, which we will see is verified
by computer simulations. Fourier inversion of Eq. (\ref{H2}), division by $\rho$, and use of (\ref{rho-chi}) yields a corresponding
low-$\chi$ expansion of the total correlation function $h(r)$ 
through second order in $\chi$ and hence has an error term of order $\chi^3$.

To get an idea of the large-$r$ asymptotic behavior of the  pair correlations, consider the 
limit $\chi \rightarrow 0$ for any $d$. In this limit, the total correlation function $h(r)$
for any $r$ obtained from Eq. (\ref{H2}) is given by
\begin{equation}
\rho h(r) = -\left(\frac{K}{2 \pi r}\right)^{d/2}J_{d/2}(Kr) \qquad (\chi \rightarrow 0),
 \label{eq_hr}
 \end{equation}
which for large $r$ is given asymptotically by 
\begin{equation}
\rho h(r) \sim -\frac{1}{r^{(d+1)/2}}\cos(r-(d+1)\pi/4) \qquad (r \rightarrow +\infty).
\end{equation}
Thus, the longed-ranged oscillations of  $h(r)$ 
are controlled by the power law  $-1/r^{(d+1)/2}$.
Equation~\eqref{eq_hr} indicates that in the limits $\chi \to 0$ and $\rho \to \infty$, $h(r) \to 0$, and therefore, the pair correlation function tends to the ideal gas even though the structure factor [Eq.~\eqref{eq_sk}] cannot tend to the ideal-gas form because of its stealthy property.
This result is in contrast to the situation considered in Fig.~\ref{rho_chi}, where we take the $\rho \to \infty$ limit by fixing $N$ and letting $v_F \to 0$. 
In that case, both $g_2(r)$ and $S(k)$ tend to the associated ideal-gas forms, i.e., $g_2(r)=1$ for all $r$ and $S(k)=1$ for all $k$.
\vspace{-0.1in}

\subsection{Comparison of theoretical predictions to simulations}

In order to test our theoretical results for the pair statistics
of stealthy ground states in the canonical ensemble, we have carried out computer simulations
to generate and sample such configurations, the details of which are described in Appendix B.
In all cases, we take $K=1$, which sets the length scale.
Our simulation results reveal that the functional trends for $S(k)$ and $g_2(r)$ predicted by the
ansatz of pseudo hard spheres in Fourier space 
with an effective packing fraction $\chi$ are remarkably accurate for a moderate range of $\chi$
about $\chi=0$.  Because it is theoretically
highly challenging to ascertain the proportionality constant $b(d)$ in Eq. (\ref{b}) that arises
in Eq. (\ref{H2}), we must rely on the simulations to guide us in its determination. First, we observe that for $d=1$, the mapping 
between $\chi$ and $\eta$ is one to one, i.e., $b(1)=1$. Second, the simulation data suggest
that, to an excellent approximation, $b(d)$ for $d\ge 2$ is given by assuming that
the peak value  of $S(k)$ or $\rho {\tilde h}(k)$, achieved at $k=K$ for sufficiently small $\chi$,
is invariant with respect to this peak value as in the  one-dimensional case, and consequently $b(d)=[\alpha(K;K) 2^d]^{-1}$.
\vspace{-0.15in}

\onecolumngrid

\begin{figure}[H]
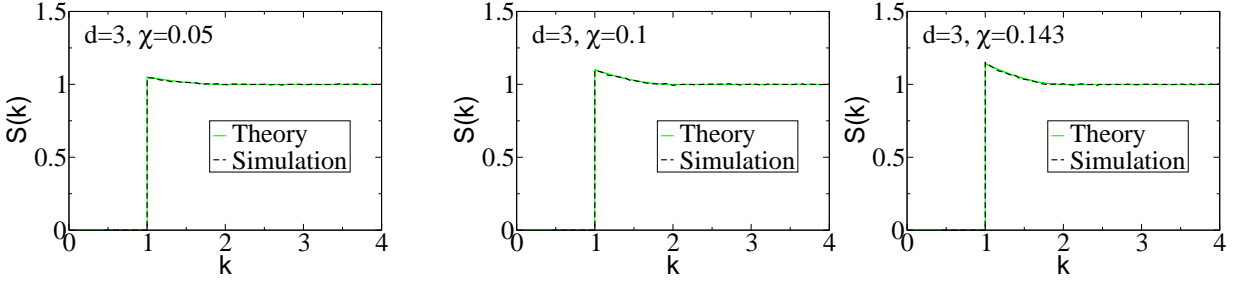

\begin{center}\includegraphics[  width=2in,
  keepaspectratio,clip=]{fig6a.eps}\hspace{0.3in}
\includegraphics[  width=2in,
  keepaspectratio,clip=]{fig6b.eps}
\includegraphics[  width=2in,
  keepaspectratio,clip=]{fig6c.eps}
\caption{Comparison of theoretical and simulation results for the structure factor 
$S(k)$ for $\chi=0.05$, 0.1, and 0.143 for $d=3$. Here, $K=1$.}\label{Sk}
\end{center}
\end{figure}
\vspace{-0.1in}
\begin{figure}[H]
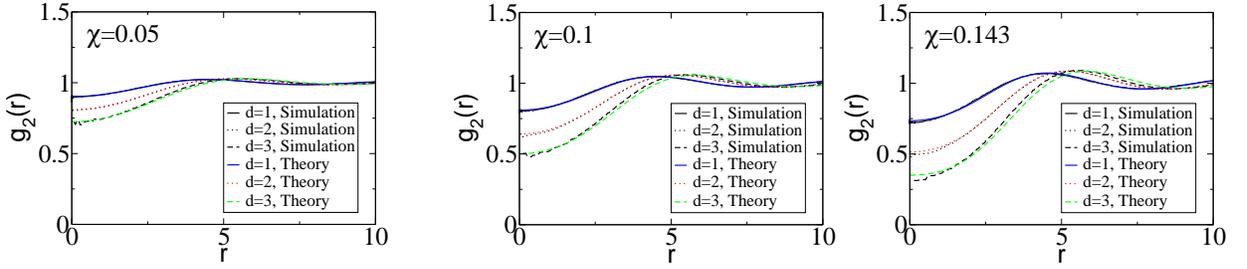

\begin{center}\includegraphics[  width=2in,
  keepaspectratio,clip=]{fig7a.eps}\hspace{0.3in}
\includegraphics[  width=2in,
  keepaspectratio,clip=]{fig7b.eps}
\includegraphics[  width=2in,
  keepaspectratio,clip=]{fig7c.eps}
\caption{Comparison of theoretical and simulation  results for the pair correlation function $g_2(r)$ for 
$\chi=0.05$, 0.1, and 0.143  across the first three space dimensions. Here, $K=1$.}\label{g-1}
\end{center}
\end{figure}
\twocolumngrid

\begin{figure}[H]
\begin{center}\includegraphics[  width=2.2in,
  keepaspectratio,clip=]{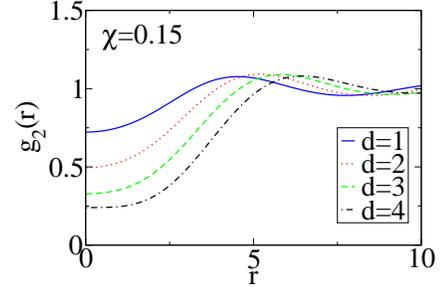}
\caption{Theoretical predictions for the pair correlation function $g_2(r)$ for 
 $\chi=0.15$ across the first four space dimensions. Here $K=1$.}\label{g-2}
\end{center}
\end{figure}

Figure \ref{Sk} shows that the structure factor $S(k)$, as obtained from Eqs. (\ref{H1}) and (\ref{H2}), 
is in excellent agreement with the corresponding simulated quantities for $\chi=0.05$, 0.1, and $0.143$ for $d=3$.
In Fig. \ref{g-1}, we compare our theoretical results for the pair correlation function $g_2(r)$,
as obtained by Fourier inversion of Eq. (\ref{H2}), to corresponding  simulation  results
across the first three space dimensions. Again, we see excellent agreement between theory and simulations,
which validates the pseudo-hard-sphere Fourier-space ansatz.
Figure \ref{g-2} depicts our theoretical predictions for $g_2(r)$ for 
 $\chi=0.15$ across the first four space dimensions. It is seen that increasing dimensionality
increases short-range correlations.

\subsection{Translational order or disorder metric}

We have seen that both short- and long-scale correlations increase as $\chi$ increases.
A useful scalar positive order metric that captures the degree to which translational order
increases with $\chi$ is given by
\begin{eqnarray}
\tau &\equiv & \frac{1}{D^d}\int_{\mathbb{R}^d} h^2(r) d{\bf r} \nonumber \\
&=& \frac{1}{(2\pi)^d D^d} \int_{\mathbb{R}^d} {\tilde h}^2(k) d{\bf k},
\end{eqnarray}
where we have used Parseval's theorem and $D$ is some characteristic
length scale \cite{Note11}.
Note that for an ideal gas (spatially uncorrelated Poisson point process), $\tau=0$
because $h(r)=0$ for all $r$. Thus, a  deviation of $\tau$ from zero
measures translational order with respect  to the fully uncorrelated
case. Because $\tau$ diverges for any perfect crystal, it is a quantity
that is better suited to distinguish the degree of pair correlations
in amorphous systems.

In the case of stealthy ground-state configurations, $\tau$ is given explicitly by
the relation
\begin{equation}
\tau=\frac{1}{(2\pi)^d \rho^2 D^d} \int_{\mathbb{R}^d} {\tilde H}^2(k) d{\bf k},
\label{t}
\end{equation}
where ${\tilde H}(k)$ is given by Eq. (\ref{H1}).
Substitution of the leading-order term in the $\chi$ expansion Eq. (\ref{H2}) 
into (\ref{t}) yields
\begin{equation}
\tau = \frac{4 d^2(2\pi)^d}{v_1(1)} \chi^2 + {\cal O}(\chi^3),
\end{equation}
where we have taken $D=K^{-1}$. Thus, for stealthy ground states, the order metric
$\tau$ grows quadratically with $\chi$ for small $\chi$. Since the error is of
order $\chi^3$, we expect that this quadratic form will be a very good approximation
of $\tau$ up to moderately large values of $\chi$. Indeed, this is confirmed by our simulations
up to $\chi=0.25$. Note that because stealthy disordered ground states 
(for sufficiently small $\chi$) are pseudo-equilibrium hard-sphere systems in Fourier space,
the form of the order metric $\tau$ [Eq. (\ref{t})] ensures that it
will behave similarly to $\tau$ for equilibrium hard spheres in direct
space for low densities.

\section{Local Number Variance for Stealthy Hyperuniform Disordered Ground States}
\label{var}

Here we investigate theoretically the local number variance for stealthy disordered
ground states as a function of $\chi$ and then use these results to extract an order metric \cite{To03a}
that describes the extent to which large-scale density fluctuations are suppressed
as $\chi$ increases in these hyperuniform systems (see Sec. \ref{hyp}).
The local number variance $\sigma^2(R)$ associated with a general statistically homogeneous
and isotropic point process in $\mathbb{R}^d$ at number density $\rho$
for a spherical window of radius $R$ is determined entirely by pair correlations \cite{To03a}:
\begin{eqnarray}
\sigma^2(R)&=&
 \rho v_1(R)\Big[ 1+\rho\int_{\mathbb{R}^d}  h(r)
\alpha(r;R) d{\bf r}\Big] \nonumber \\
&=&\rho v_1(R)\Big[\frac{1}{(2\pi)^d} \int_{\mathbb{R}^d} S(k) 
{\tilde \alpha}(k;R) d{\bf k}\Big]  ,
\label{local}
\end{eqnarray}
where  $v_1(R)$ is the $d$-dimensional volume of a spherical window [cf. Eq. (\ref{vol})], $h(r)$
is the total correlation function [cf. Eq. (\ref{tot})],
$\alpha(r;R)$ is the scaled intersection volume of two spherical windows
of radius $R$, as given by Eq. (\ref{alpha}), and ${\tilde \alpha}(k;R)$ is the Fourier transform
of $\alpha(r;R)$, which is explicitly given by \cite{To03a}
\begin{equation}
{\tilde \alpha}(k;R)= 2^d \pi^{d/2} \Gamma(1+d/2)\frac{[J_{d/2}(kR)]^2}{k^d}.
\end{equation}

We have already noted that the stealthy ground states considered in the
present paper are hyperuniform, i.e., $S(k) \rightarrow 0$ as $k \rightarrow 0$ (see Sec. \ref{hyp}). This means that such systems
obey the sum rule $\rho \int_{\mathbb{R}^d} h(r) d{\bf r}=-1$ and, because of the rapid manner
in which $S(k)$ vanishes in the limit $k \rightarrow 0$, the number variance
has the following large-$R$ asymptotic behavior \cite{To03a}:
\begin{equation}
\sigma^2(R) = \Lambda(R) R^{d-1} + {\cal O}(R^{d-3}),
\label{scaling}
\end{equation}
where $\Lambda(R)$ is a bounded function that oscillates around an average value
\begin{equation}
{\overline \Lambda} = \lim_{L \rightarrow \infty} \frac{1}{L} \int_0^L \Lambda(R) dR.
\end{equation}
The scaling (\ref{scaling}) occurs for a broader class of hyperuniform
systems, as specified by relation (\ref{NVscaling}). The parameter $\overline \Lambda$
is an order metric that quantifies the extent to which large-scale
density fluctuations are suppressed in  such hyperuniform systems \cite{To03a}. 
To compare different hyperuniform systems, Torquato and Stillinger 
used the following rescaled order metric:
\begin{equation}
{\overline B}= \frac{\overline \Lambda}{\phi^{(d-1)/d}},
\label{metric}
\end{equation}
which is independent of the density, where $\phi=\rho v_1(1/2)$.
Among all hyperuniform point patterns having the scaling (\ref{scaling}), 
${\overline B}$ is minimized (greatest suppression of large-scale density fluctuations)
for the integer, triangular,  BCC, and $D_4$ lattices for $d=1$, 2, 3, and 4, respectively \cite{To03a,Za09}.

\begin{figure}
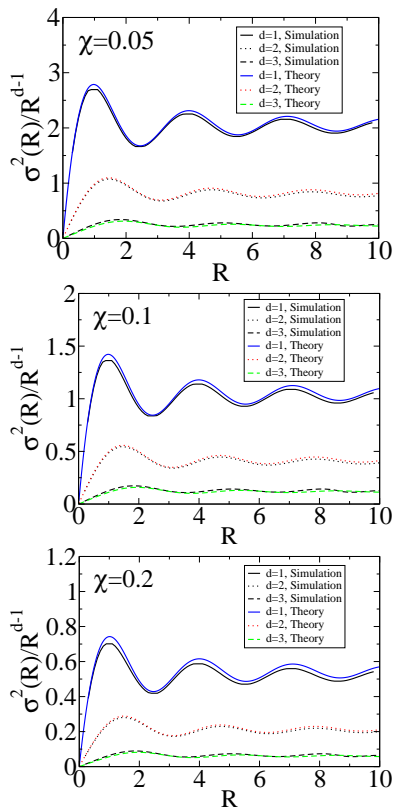

\begin{center}\includegraphics[  width=2.in,
  keepaspectratio,clip=]{fig9a.eps}\\\includegraphics[  width=2.in,
  keepaspectratio,clip=]{fig9b.eps}\\ \includegraphics[  width=2.in,
  keepaspectratio,clip=]{fig9c.eps}
\caption{Comparison of analytical and numerical results for the number variance for $\chi=0.05$, 0.1, and $0.2$
across the first three space dimensions. Here, $K=1$.}
\label{number}
\end{center}
\end{figure}

Using the analytical results for $g_2(r)$ or $S(k)$ described in the previous section, 
we have computed relation (\ref{local}) for $\sigma^2(R)$ versus $R$ for selected values of $\chi$ across the first 
three dimensions
and compared them to our corresponding simulation results; see Fig. \ref{number}.
The analytical and numerical results are in excellent
agreement with one another. Table \ref{table-var} lists the order metric $\overline B$ 
for various values of $\chi$ across the first four dimensions for disordered stealthy ground states, 
as obtained from the analytical estimates
of $\sigma^2(R)$ and Eq. (\ref{metric}). 
These results  are also compared to the corresponding optimal
values. As expected, $\overline B$ decreases as $\chi$ increases
for fixed $d$.

\begin{table}[H]
\caption{The order metric $\overline B$ for various values
of $\chi$ across the  first four dimensions  for disordered stealthy ground states
for selected values of $\chi$ up to $\chi=0.25$, as obtained from the analytical estimates
of $\sigma^2(R)$ and Eq. (\ref{metric}). Included for comparison
are the structures in each dimension that have the minimal values of $\overline B$,
all of which are Bravais lattices \cite{To03a,Za09}.}
\begin{center}
\begin{tabular}{|c|c|c|c|c|c|}
$\chi$ & $d=1$ & $d=2$ & $d=3$ &$d=4$\\ \hline

0.05 & 2.071 & 1.452& 2.164& 4.560\\

0.1 & 1.051& 1.040& 1.738& 3.875\\

0.143 & 0.745& 0.880& 1.558& 3.576\\

0.2 & 0.54& 0.755& 1.411& 3.327\\

0.25 & 0.439& 0.683& 1.325& 3.179\\

\hline
Integer lattice &0.167 & & &\\
Triangular lattice & & 0.508 & &\\
BCC lattice & & & 1.245 & \\
$D_4$ lattice & & & & 2.798 \\
\end{tabular}
\end{center}
\label{table-var}
\end{table}

\section{Nearest-Neighbor Functions}
\label{near}

Here, we obtain theoretical predictions for the nearest-neighbor functions 
of stealthy disordered ground states. Nearest-neighbor functions describe the probability of finding
the nearest point of a point process in $\mathbb{R}^d$  at some given distance from a
reference point in space. Such statistical quantities are
called ``void" or ``particle" nearest-neighbor functions
if the reference point is an arbitrary
point of space or an actual point of the point process, respectively \cite{To90c}.
Our focus here is on the particle nearest-neighbor functions.

The particle nearest-neighbor probability density function $H_{P}(r)$ is defined such that
$H_{P}(r) dr$ gives the probability that the nearest point to the arbitrarily chosen point
lies at a distance between $r$ and $r + dr$ from this chosen
point of the point process. The probability that a sphere of radius $r$ centered
at a point does not have other points, called the exclusion probability $E_P(r)$, is the
associated complementary cumulative distribution
function and so
$E_{P}(r) = 1 - \int_{0}^{r} H_{P}(x) \, dx$ and
and hence $H_{P}(r) = -\partial E_{P}/\partial r$.

The nearest-neighbor functions can be expressed as an infinite series
whose terms are integrals over $n$-body correlation functions defined in Sec. \ref{def} \cite{To90c,To02a}.
In general, an exact evaluation of this infinite series is not possible because the $g_n$ are not known
accurately for $n \ge 3$, except for
simple cases, such as the Poisson point process. Theoretically, one must
either devise approximations or rigorous bounds to estimate
nearest-neighbor quantities for general models \cite{To90c,To95a}.

Torquato has given rigorous upper and lower bounds on
 the so-called  {\it canonical $n$-point correlation function} $H_n$ for
point processes in $\mathbb{R}^d$ \cite{To86i,To02a}.  Since nearest-neighbor functions are just special cases of $H_{n}$,
 then we also have strict bounds on them for such models \cite{To90c,To02a}.
Here, we employ upper and lower bounds on $E_P(r)$, which
relies on knowledge of the pair correlation function:
\begin{eqnarray}
E_P(r) \ge 1 - Z(r), \label{EP-lower}\\
E_P(r) \le \exp[-Z(r)], \label{EP-upper}
\end{eqnarray}
where $Z(r)$ is the {\it cumulative coordination number} [cf. Eq. (\ref{cum})]. 
The upper bound (\ref{EP-upper}) was presented in Ref. \cite{To08c}.

\begin{figure}[H]
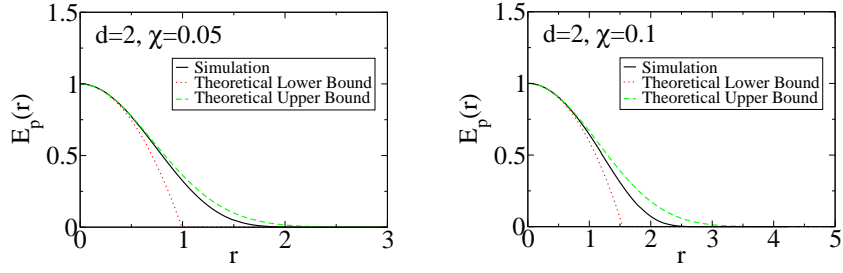

\begin{center}\includegraphics[  width=2in,
  keepaspectratio,clip=]{fig10a.eps}\hspace{0.3in}
\includegraphics[  width=2in,
  keepaspectratio,clip=]{fig10b.eps}
\caption{Comparison of  the lower and upper bounds (\ref{EP-lower}) and (\ref{EP-upper})  
to our numerical results for the exclusion probability function $E_P(r)$ for $\chi=0.05$
and $\chi= 0.1$ for $d=2$. Here, $K=1$.}
\label{EP-2}
\end{center}
\end{figure}

\begin{figure}[H]
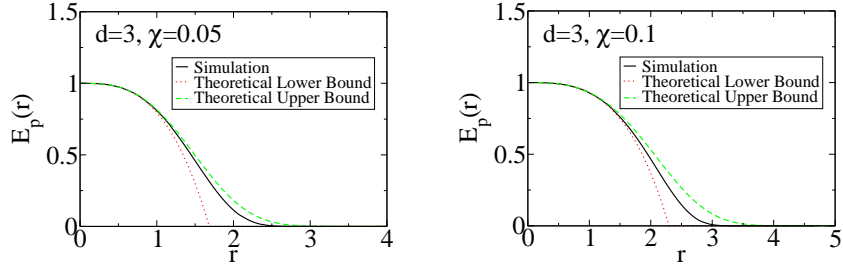

\begin{center}\includegraphics[  width=2in,
  keepaspectratio,clip=]{fig11a.eps}\hspace{0.3in}
\includegraphics[  width=2in,
  keepaspectratio,clip=]{fig11b.eps}
\caption{Comparison of  the lower and upper bounds (\ref{EP-lower}) and (\ref{EP-upper}) 
to our numerical results for the exclusion probability function $E_P(r)$ for $\chi=0.05$
and $\chi= 0.1$ for $d=3$. Here, $K=1$.}
\label{EP-3}
\end{center}
\end{figure}

These bounds are evaluated for stealthy ground-state configurations using the analytical expression
for the pair correlation function given in Sec. \ref{pseudo}.
Figures \ref{EP-2} and \ref{EP-3} compare these bounds to our numerical
results for both $\chi=0.05$ and $\chi=0.1$ for $d=2$ and $d=3$, respectively.
We see that the bounds on $E_P(r)$ provide the correct qualitative trends 
as a function of $r$; the upper bound being the sharper of the two bounds for these cases.

\onecolumngrid

\twocolumngrid

The mean nearest-neighbor distance $\lambda$ is defined as the first moment of $H_P(r)$ or, equivalently,
zeroth moment of $E_P(r)$, i.e.,
\begin{equation}
\lambda = \int_0^\infty r H_P(r) dr = \int_0^\infty  E_P(r) dr.
\label{mean-1}
\end{equation}
For an ideal gas (Poisson point process) at number density $\rho$, the mean nearest neighbor
can be explicitly given in any dimension \cite{To02a}:
\begin{equation}
\lambda_{Ideal} =\frac{\Gamma(1+1/d)}{2\,[\rho v_1(1/2)]^{1/d}}.
\label{mean-2}
\end{equation}

\begin{figure}[H]
\begin{center}
\includegraphics[  width=2in,keepaspectratio,clip=]{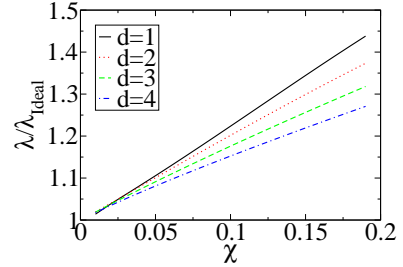}
\caption{Upper bounds on the mean nearest-neighbor distance $\lambda$, scaled by $\lambda_{ideal}$
[cf. Eq. (\ref{mean-2})], as a function of $\chi$ for the first four space dimensions. }
\label{lambda}
\end{center}
\end{figure}

Using the upper bound (\ref{EP-upper}) and relation (\ref{mean-1}), we plot in Fig. \ref{lambda} 
upper bounds on the mean nearest-neighbor distance $\lambda$, scaled by the corresponding ideal-gas 
quantity obtained from Eq. (\ref{mean-2}), as a function of $\chi$ for the first four space dimensions.
For fixed $\chi$, the upper bounds on $\lambda/\lambda_{ideal}$ decrease
as the space dimension increases, as expected, and tends to unity in the large-$d$
limit, consistent with the so-called ``decorrelation" principle \cite{To06b,To08c}.

\section{Excited States: Structure Factor and Thermal Expansion Coefficient}
\label{excite}

Here, we derive accurate analytical formulas for the structure factor 
and thermal expansion coefficient for the excited states associated with stealthy ground states at sufficiently small temperatures.
We see from the compressibility relation (\ref{comp}) that if the isothermal compressibility $\kappa_T$
is bounded, then $S(0)$ must be zero for any ground state, stealthy or not. Recall that 
for stealthy ground states, $\kappa_T$ is bounded according to Eq. (\ref{iso}).
Now consider excited states infinitesimally close to the
stealthy ground states, i.e., when temperature $T$ is positive 
and infinitesimally small. Under the highly plausible assumption that the
structure of such excited states will be infinitesimally near
the ground-state configurations for sufficiently small $\chi$ and $T$, then to an excellent approximation,
the pressure is given by
\begin{equation}
p \sim \rho T+ \frac{\rho^2}{2},
\label{p-approx}
\end{equation}
where the first term is the ideal-gas contribution and the second term is the configurational
contribution, which, under the stated conditions, is effectively the same as the ground-state
expression (\ref{p-u}), where we have set  $k_B=v_0=K=1$. Thus, relation (\ref{p-approx}) yields the isothermal compressibility 
$\kappa_T=[\rho(\rho+T)]^{-1}$, which, when substituted into Eq. (\ref{comp}) for large $\rho$ (small $\chi$)
and small $T$, yields that $S(0)$ varies linearly with $T$ for such excited states:
\begin{equation}
S(0) \sim C(d) \,\chi\, T,
\label{S0}
\end{equation}
in units $k_B=v_0=K=1$, where $C(d)= 2d \,(2\pi)^d /v_1(1)$ is a $d$-dependent constant.

\begin{figure}[H]
\begin{center}\includegraphics[  width=2.0in,keepaspectratio,clip=]{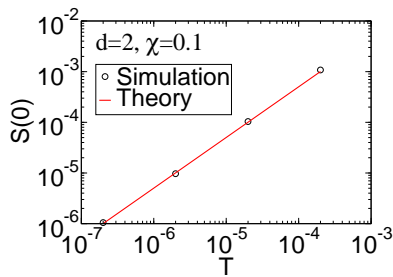}
\caption{Comparison of the theoretically predicted structure factor at the origin $S(0)$ versus absolute temperature $T$,
as obtained from Eq. (\ref{S0}), to our corresponding simulation results for
a two-dimensional stealthy system at $\chi=0.1$. Here we take $k_B=v_0=K=1$.}\label{ex}
\end{center}
\end{figure}

Figure \ref{ex} shows that the prediction of relation (\ref{S0})
is in excellent agreement with  our MD simulation results (Appendix B) in the case $d=2$.
It is expected that this positive value of $S(0)$ will be the uniform 
value of $S(k)$ for $0 \le k  \le K$ for the special case of the step-function
power-law potential ${\tilde v}(K)$ [the case $m=0$ in  Eq. (\ref{power})] for small $\chi$.
This behavior of $S(k)$ has indeed been verified by 
our simulation results in various dimensions.
For other stealthy-potential function choices, $S(k)$ will no longer be a constant for $0 \le k  \le K$.

An interesting conclusion to be drawn from this analysis 
is that, for a system, $S(0)$ can 
be arbitrarily close to zero at positive temperatures, even if $T$ is itself arbitrarily small.
This means that, for all practical purposes, such systems at positive $T$ are effectively hyperuniform.
Perfect hyperuniformity is not necessarily required in order to achieve
novel physical properties in technological applications.

Using the cyclic identity $(\partial \rho/\partial T)_p (\partial T/\partial p)_\rho  (\partial p/\partial \rho)_T=-1$
and approximation (\ref{p-approx}), it immediately follows that the thermal expansion coefficient $\alpha \equiv -\rho^{-1} 
(\partial \rho/\partial T)_p$, for sufficiently small $\chi$ and $T$, is given by 
\begin{equation}
\alpha \sim C^2(d) \chi^2 T.
\end{equation}
We see that the thermal expansion is positive under such conditions, which is to be contrasted
with the anomalous negative thermal expansion behavior for sufficiently large $\chi$
over a low temperature range demonstrated in our earlier numerical work \cite{Ba09a,Ba09b}.

\section{Conclusions and Discussion}
\label{concl}

Stealthy hyperuniform disordered ground states in $\mathbb{R}^d$ are infinitely degenerate
and arise from a class of bounded long-ranged pair potentials with compactly supported Fourier transforms. 
Such exotic many-particle states of matter were 
previously studied only numerically. Because the configurational dimensionality  depends on the density (or $\chi$), a highly unusual situation,
and there are an infinite number of distinct ways to sample the
ground-state manifold,  each with its own probability measure, 
it has been theoretically very challenging to devise predictive ensemble
theories. A new type of statistical-mechanical theory needed to be  invented.
This paper has initiated such a theoretical program. 

Specifically, we
have  derived general exact relations for the ground-state energy, pressure, and isothermal compressibility
that apply to any  ensemble as a function of the number density $\rho$ in any dimension $d$.
We demonstrated  how  disordered degenerate ground states can arise as part of
the ground-state manifold. We also obtained  exact integral conditions that  both the pair correlation function 
$g_2(r)$ and structure factor $S(k)$ must satisfy in any ensemble. Then, we  specialized our results to the canonical ensemble 
in the zero-temperature limit by exploiting an ansatz that stealthy states behave 
like pseudo-equilibrium  hard-sphere systems in Fourier space \cite{Note13}.
The resulting theoretical predictions for $g_2(r)$ and $S(k)$ were shown to be in excellent
agreement with computer simulations across the first three space dimensions for sufficiently
small $\chi$. These results were used to theoretically obtain  order metrics, local number variance, and nearest-neighbor
functions across dimensions.  We also derived accurate analytical formulas for the structure factor and thermal expansion coefficient 
for the excited states associated with stealthy ground states at sufficiently small temperatures.
Our analyses provide new insights on our fundamental understanding of the nature and formation of
low-temperature states of amorphous matter. Our work also offers challenges to experimentalists
to synthesize stealthy ground states at the molecular level, perhaps
with polymers, as suggested in Sec. \ref{family}.

There are many remaining open theoretical problems.
While the pseudo-hard-sphere system picture for the canonical
ensemble is almost surely exact in the limit $\chi \rightarrow 0$, a 
future challenge would be to provide rigorous justification for this picture for positive but small $\chi$. 
One possible avenue that could be pursued  is the formulation of an exact perturbation theory for the pair statistics
about the weakly constrained limit ($\chi \rightarrow 0$).
As noted in the Introduction, while 
the configuration space is fully connected for sufficiently small $\chi$,
quantifying its topology as a function of $\chi$ up to  $\chi_{max}^*$ is an outstanding 
open problem. At some intermediate range of $\chi$, the topology of the ground-state 
manifold undergoes a sequence of one or more disconnection events, but this process
is poorly understood and demands future study. In the limit $\chi \rightarrow \chi_{max}^*$, the 
disconnection becomes complete at the unique crystal ground state \cite{Note12}.

\begin{figure}[H]
\begin{center}
\includegraphics[  width=3in,  keepaspectratio,clip=]{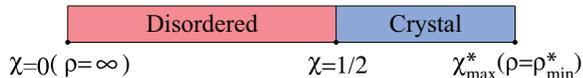}
\caption{Phase diagram for the entropically favored stealthy ground states in the canonical ensemble
as a function of $\chi$ which applies to the first four space dimensions.  }\label{phase}
\end{center}
\end{figure}

The simple constraint and degrees-of-freedom counting arguments described earlier lead
to definite predictions for the entropically favored stealthy ground states
derived from the canonical ensemble in the limit $T \rightarrow 0$.
For $\chi$ between 0 and 1/2, the ground states are disordered and possess
a configurational dimension per particle $d_C$ of $d(1-2\chi)$; see Eq. (\ref{dc}). At $\chi=1/2$, 
the configurational dimensionality per particle collapses to zero, and there is a 
concomitant  phase transition to a crystal phase. The fact that  the pressure is a continuous function
of density [cf. Eq. (\ref{p-u})] for all $\chi$ up to $\chi_{max}^*$ (see Tables I-IV) implies
that any phase transition  could  be continuous. While this eliminates a first-order phase transition in which the
phase densities are unequal, it does not prohibit such a phase transition
in which the two distinct phases possess the same number density. At $\chi=0.5$, 
our numerical evidence indicates that the structurally distinct fluid and crystal phases have equal free energies.
This does not necessarily imply that the two phases could coexist side by side within the system separated by an interface. 
This phase diagram is depicted
in Fig. \ref{phase}, which applies to the first four space dimensions.
For $d=1$, the only crystal phase allowed is the integer lattice (see Sec. \ref{exist}),
and hence there can be no phase coexistence. 
For $d=2$, our simulations \cite{Ba08,Ba09a,Ba09b,Zh14} indicate that the crystal
phase is the triangular lattice for $1/2 \le \chi \le \chi_{max}^*$. However, for $d=3$,
it is possible that there may be more than one crystal phase. For example, while for sufficiently
high $\chi$ up to $\chi_{max}^*$, we expect the stable crystal to be the BCC lattice, our simulations cannot eliminate
the possibility that the FCC lattice is a stable phase for some $\chi$ in the range
$1/2 \le \chi \le 0.9068\ldots$; for $\chi> 0.9068\ldots$, the FCC lattice cannot be a ground
state (see Table III). Since four dimensions is more similar to two dimensions in that
the lattice corresponding to  $\chi_{max}^*$ is equivalent to its dual, we would expect
that the $D_4$ lattice is the stable crystal for $1/2 \le \chi \le \chi_{max}^*$, but this
remains to be confirmed.

All of our previous  and current simulations for the first three space
dimensions \cite{Uc04b,Uc06b,Ba08,Ba09a,Ba09b} strongly suggest that all of the energy minima attained were global ones for $\chi<0.5$, 
but  when $\chi>0.5$, the topography of the energy landscape suddenly
exhibits local minima above the ground-state energies. The possible configurations that can arise
as part of the ground-state manifold for $\chi > 1/2$, regardless
of their probability of occurrence, not only include periodic crystals for $d\ge 2$, as discussed
in Sec. \ref{exist}, but also aperiodic structures, reflecting the complex
nature of the energy landscape. For example, for $d=2$,
the manifold includes aperiodic ``wavy phases," which have been shown
to arise via numerical energy minimizations from random initial
conditions with high probability in a range of $\chi$ where the triangular
lattice is entropically favored \cite{Uc04b,Ba08}. A deeper understanding 
of such aspects of the ground-state manifold would undoubtedly
shed light on the topography of the energy landscape.

\acknowledgments{This research was supported by the US
Department of Energy, Office of Basic Energy Sciences,
Division of Materials Sciences and Engineering, under Award
No. DE-FG02-04-ER46108.}

\appendix

\noindent
 \section{{COMMON $d$-DIMENSIONAL LATTICES}}
\label{app_lattices}
Common $d$-dimensional lattices include the hypercubic {$\mathbb{Z}^d$,
}{checkerboard} {$D_d$, and }{root} {$A_d$ lattices,
defined, respectively, by
}\begin{equation}{
\mathbb{Z}^d=\{(x_1,\ldots,x_d): x_i \in }{{\mathbb{ Z}}}{\} \quad \mbox{for}\; d\ge 1
}\end{equation}
\begin{equation}{
D_d=\{(x_1,\ldots,x_d)\in \mathbb{Z}^d: x_1+ \cdots +x_d ~~\mbox{even}\} \quad \mbox{for}\; d\ge 3
}\end{equation}
\begin{eqnarray}
A_d&=&\{(x_0,x_1,\ldots,x_d)\in \mathbb{Z}^{d+1}: x_0+ x_1+ \cdots +x_d =0\} \nonumber \\
&& \quad \mbox{for}\; d\ge 1,
\end{eqnarray}
{where $\mathbb{Z}$ is the set of integers ($\ldots -3,-2,-1,0,1,2,3\ldots$);
$x_1,\ldots,x_d$ denote the components of a lattice vector
of either $\mathbb{Z}^d$ or $D_d$; and $x_0,x_1,\ldots,x_d$ denote
a lattice vector of $A_d$. The $d$-dimensional lattices $\mathbb{Z}^d_*$, $D_d^*$ and $A_d^*$ are
the corresponding dual lattices. 
Following Conway and Sloane \mbox{%DIFAUXCMD
\cite{Co93}
}%DIFAUXCMD
, we say that
two lattices are }{\it {equivalent}} {or }{\it {similar}} {if one becomes identical
to the other possibly by a rotation, reflection, and change of scale,
for which we use the symbol $\equiv$.  
The $A_d$ and $D_d$ lattices can be regarded as $d$-dimensional generalizations
of the face-centered-cubic (FCC) lattice defined by $A_3 \equiv D_3$; however, for $d\ge 4$, they are no longer 
equivalent. In two dimensions, $A_2 \equiv A_2^*$ defines the triangular lattice
with a dual lattice that is equivalent.
In three dimensions, $A_3^* \equiv D_3^*$ defines the body-centered-cubic (BCC)
lattice. 
%DIF > A {\it self-dual} lattice $\Lambda$ is one with an {\it identical} dual lattice $\Lambda^*$ at density $\rho = \rho_* = 1/(2\pi)^{d/2}$, i.e., without any rotation,
%DIF > reflection, or change of scale for which we write $\Lambda=\Lambda^*$.\footnote{
%DIF > Mathematicians usually define a dual Bravais lattice to have a
%DIF > fundamental cell volume $v_{F^*} = 1/v_{F}$ [i.e., without the factor of
%DIF > $(2\pi)^d$], in which case self-duality is defined with respect to unit density; see Ref.~\cite{Co93}.}
In four dimensions, the checkerboard lattice and its dual are equivalent,
i.e., $D_4\equiv D_4^*$. The hypercubic lattice $\mathbb{Z}^d\equiv \mathbb{Z}^d_*$ and its dual
lattice are equivalent for all $d$.
}

 \section{SIMULATION PROCEDURE TO GENERATE AND SAMPLE STEALTHY GROUND STATES IN THE CANONICAL ENSEMBLE}

To numerically sample the stealthy ground-state manifold in the disordered regime in 
the  canonical ensemble in the $T \to 0$ limit for $d=1$, 2, and 3, we performed molecular dynamics (MD) simulations at a very low dimensionless
equilibration temperature $T_E\equiv k_B T/(v_0 K^d)$, periodically took configurational ``snapshots,'' 
and then used these configurations as input to the L-BFGS optimization algorithm \cite{Liu89} 
to get the corresponding  ground states. The dimensionless temperatures that we use are 
$T_E=2\times 10^{-4}$ for $d=1$, $T_E=2\times 10^{-6}$ for $d=2$, and $T_E=1\times 10^{-6}$ for $d=3$.
The equilibration temperature at a fixed dimension was chosen so that no changes in the pair
correlation function are observed over some range of equilibration temperatures. 
The MD simulations were first performed in the microcanonical ensemble using the velocity 
Verlet algorithm \cite{Fr96}. The time steps were chosen so that the relative
energy change every 3000 time steps is less than $10^{-8}$.  However, to enforce the desired temperature, 
we also performed MD simulations in the canonical ensemble using an Anderson thermostat
\cite{Fr96}. We employed fifteen million time steps to equilibrate a system. After that, 
a snapshot was taken every 3000 time steps for further energy minimization.
Because stealthy potentials in direct space are long-ranged, the energy is most accurately
calculated in Fourier space using Eq. (1).
% (which is equivalent to summing over all periodic images in the directspace in the infinite-volume limit). 
The numerical errors in the achieved ground-state energies are extremely small, usually on the order of $10^{-20}$  (in units of $v_0 K^d$). The force on the $j$th particle is calculated using the gradient of Eq. (1), yielding
${\bf F}_j=-{\bf \bigtriangledown}_j \Psi({\bf r}^N)=\frac{1}{v_F}\sum_{\bf k} {\bf k}\, {\tilde v}({\bf k})\,\mbox{Im}[
{\tilde n}({\bf k})\exp(i {\bf k\cdot r}_j)] $.
The number of particles in the simulation box, $N$, is calculated from Eq. (2) for given $d$, $\chi$ and $M(K)$. 
We chose $M(K)=50$ for $d=1$, $M(K)=54$ for $d=2$, and $M(K)=39$ for $d=3$. The fundamental
cell employed is the one corresponding to the crystal with the largest value of
$\chi_{max}$ in each dimension (see Tables I-IV).

Structural characteristics, such as the pair correlation function $g_2(r)$, structure factor
$S(k)$, number variance $\sigma^2(R)$ and nearest-neighbor function $E_P(r)$, by sampling each generated configuration
for a fixed value of $\chi$ and ensemble averaging over at least 20000 configurations.
Two power-law potentials (\ref{power}) were used: one with $m=0$ and the
other with $m=2$. As expected, both potentials produced the same ensemble-averaged
structural properties to within small numerical errors (as explained in Sec. \ref{dim}), the agreement of 
which provides a good test on the validity of the simulation results. Our simulation results for $g_2(r)$ and 
$S(k)$ also satisfied the exact integral conditions presented in Secs. \ref{pair-1} and \ref{pair-2}. 
Additional simulation details will be described elsewhere \cite{Zh14}.

%\bibliographystyle{prsty}
%\bibliography{new}

\begin{thebibliography}{78}
\bibitem{X}
The limit  $\rho \rightarrow \infty$ described in this caption does not commute with the limit of the number density $\rho$ going to infinity, where  $\rho$ is derived from the thermodynamic limit defined in Eq.~\eqref{u1}, which is the situation that will be employed throughout the remainder of the paper.
If we take the thermodynamic limit first and then take the limit $\rho \rightarrow \infty$, the pair correlation function tends to the trivial ideal-gas form, but the structure factor does not, as detailed in Sec.~\ref{subsec7a}.
\end{thebibliography}

\end{document}